\documentclass[review]{elsarticle}

\usepackage{lineno}
\usepackage{amssymb}
\usepackage{amsmath}
\usepackage{bm,boxedminipage,float}
\usepackage[english]{babel}
\usepackage[colorlinks,linkcolor=blue,anchorcolor=blue,citecolor=blue]{hyperref}
\newtheorem{definition}{Definition}
\newtheorem{lemma}{Lemma}
\newtheorem{theorem}{Theorem}
\newenvironment{proof}{\vspace{8pt}
\noindent{\bf Proof}: }{{\hfill {\large $\Box$}} \vspace{8pt}}

\modulolinenumbers[5]

\journal{~}

\begin{document}

\begin{frontmatter}

\title{Multi-Server Verifiable Delegation of Computations: Unconditional Security  and  Practical Efficiency}

\author{Liang Feng Zhang
\footnotetext{\em Preprint submitted to {\em Information and Computation}: \href{https://doi.org/10.1016/j.ic.2021.104740}{\color{blue}\em10.1016/j.ic.2021.104740}}}
\address{School of Information Science and Technology, \\
ShanghaiTech University, \\ Shanghai, PR China}


\ead{zhanglf@shanghaitech.edu.cn}


\begin{abstract}
Outsourcing computation has gained significant popularity in recent years
  due to the prevalence  of  cloud computing.
  There are two main   security concerns in outsourcing computation:
  how to guarantee the cloud server performs the computation correctly  and
   how to keep  the client's data secret. The {\em single-server verifiable computation}
   (SSVC)
   of Gennaro, Gentry and Parno (Crypto'10)
  enables a client to delegate the computation of a  function
$f$  on any input $x$
  with both concerns highly relieved,
  but only results in {\em computationally secure} schemes that
     {\em lack practical efficiency}.

While the SSVC schemes   use a single server, in this paper we
develop a {\em multi-server verifiable computation} (MSVC)  model where the client  shares both
$f$ and  $x$ among multiple  servers, each server performs a set of
computations on its shares, and finally the client reconstructs $f(x)$ from
all servers' results.
In this MSVC model we propose a generic construction  for outsourcing
computations of the form $F{\bf x}$, where $F$ is a matrix and $\bf x$ is a  vector.
Our generic construction achieves    {\em information-theoretic   security,
input privacy} and {\em function privacy}.
By optimizing the parameters, we obtain  both a   3-server  scheme,which uses  the least number of servers, and
a   4-server  scheme, which incurs  the least workload.
By decomposing many polynomial computations  as a two-stage computation,
  where the first-stage   has the form $F{\bf x}$ and  the second-stage   is  fast,
  and  delegating the first-stage computation, we
obtain  MSVC schemes for these  polynomials.
We  implement  our MSVC schemes and show  that they are
among the most {\em practical} ones   to date.
\end{abstract}


\end{frontmatter}

  \section{Introduction}

Outsourcing of  computation and data has gained significant popularity in recent years
  due to the prevalence of cloud computing.
The computationally weak devices  such as smartphones   can offload the storage of large-scale data
and expensive computations on the data to
  powerful  cloud services in a pay-per-use manner, which is both  scalable and economical.
There are two fundamental  security concerns in outsourcing:
  how to ensure that  the cloud server      performs
  computations correctly; and  how to keep the client's data secret.

There are many  emerging   solutions \cite{snark,mac,GGP10,qsp,sig} for verifying
 the cloud servers' work. Among them is the {\em single-server verifiable computation}
 (SSVC)  of \cite{GGP10}, which   enables  the  client to outsource the computation of a function $f$ on
 any input $x$ to a cloud server, receive both the result $y$ and a cryptographic proof from the server, and then
 verify if $y=f(x)$. The    proof is designed such that    no  malicious  server
  is able to     persuade the client  to both accept its result and output
a wrong value.
Such solutions should also satisfy certain {\em efficiency}  requirements:
the client-side computation in outsourcing   should be  substantially faster   than
the naive computation of
  $f({x})$.
Minimizing the client's computational   cost
 has been one of the main  objectives in this field.
  The function $f$ should be encoded and then given to the  server.
Preparing the encoding may be a heavy task and the SSVC  of \cite{GGP10} adopts an {\em amortized model} where
the one-time cost of encoding $f$ is amortized over the computations  of $f$ on  many different inputs
$x$.

The problem of keeping both $f$ and   $x$ secret  has also been   studied as well.
The scheme   of \cite{GGP10} attains both {\em input privacy}
and {\em function privacy} by using the very expensive   machineries such as
fully homomorphic encryption (FHE) and garbled circuits (GCs). As
 these machineries incur significant latencies in both the
  client-side computation  and the server-side computation,
   the scheme of \cite{GGP10} is rather   inefficient. Achieving
     input privacy in SSVC schemes  is
highly non-trivial. In particular, when $f$ requires high-degree computations on $x$,
the input privacy somehow requires the client to encrypt $x$ with
 a semantically secure encryption, which must allow the server to compute
$f$   on the ciphertext of $x$.
As a result, the encryption   must be   homomorphic.
However, the theoretical achievements in
 homomorphic encryption \cite{BV11,fhe} hasn't changed the situation that few
SSVC schemes based on them are practical.

\subsection{Our Work}


In this paper we try to resolve the conflicts between security/privacy and
efficiency   with a {\em multi-server verifiable computation} (MSVC) model.
In this model, we propose a generic construction for outsoucing matrix-vector multiplications  of the form
$F{\bf x}$, where $F$ is a matrix and $\bf x$ is a column vector.
Our construction is information-theoretically secure such that
the servers are not able to persuade the client to both accept the servers' results and output a wrong value. Our construction is information-theoretically input private and function private such that no single server is able learn any information about
$F$ or $\bf x$. Our construction is free of FHE, GCs, and any other public-key operations
and thus achieves practical efficiency.
By  optimizing the parameters, we get two instantiations of the generic construction:
one uses  3 servers and the other uses  4 servers.
The 3-server scheme is optimal  in terms of the total number of needed servers
and  the 4-server scheme is optimal in terms of the total workload.
We also decompose various polynomial computations into two-stage computations.
By delegating the heavier first-stage computations, which are matrix-vector multiplications, we obtain polynomial outsourcing schemes that have the same properties of security, privacy and efficiency.
 We also show applications of our schemes in
 analysis of sensitive data, polynomial outsourcing, and
  outsourced private information retrieval (PIR) \cite{HG13,MBC13}.

 \vspace{1mm}
 \noindent
 {\bf Multi-server verifiable computation.}
 In our MSVC model,  the client  shares its function
$f$ and   input $x$ among multiple   servers, each server performs a set of
computations on the function and input shares, and finally the client reconstructs $f(x)$ from
all servers' results.
An MSVC  scheme is {\em secure} if the servers cannot persuade the client to
 both accept their computation results and reconstruct a wrong function value
  $\hat{y}\neq f(x)$.
In order to
achieve the similar security properties,  the existing multi-server schemes \cite{ACG+14,CRR11}
for outsourcing computations have required  that the servers should not collude with each other.
In contrast, {\em our   schemes will be secure even if all servers are malicious and colluding with each other}.
However, when it comes to function/input privacy, we
  require that the servers should not collude with each other; otherwise, the
  servers could be able  to easily recover $f$ or $x$ from their joint shares.
The requirement of  {\em  non-colluding} servers
for input/function privacy  can be met when the servers belong to different cloud services or
  have conflicts of interest.
An MSVC scheme is {\em input private} if no
individual server is able to distinguish between its shares of two different
inputs.
An MSVC scheme is {\em  function private} if no individual server
is able to distinguish between its shares  of two different functions.
In the formal definitions of input (resp. function) privacy, it suffices to require that
each server's input (resp. function) shares are statistically  independent of
the  input (resp. function).
Our security and privacy will be information-theoretic and  requrie no  number-theoretic assumptions.

\vspace{2mm}
\noindent
{\bf Matrix outsourcing.}
 Let $\mathbb{Z}_q$ be a finite field of $q$ elements and let $m,d$ be integers.
We   interpret  any matrix $F\in \mathbb{Z}_q^{m\times d}$ as a function that takes any vector
${\bf x}\in\mathbb{F}_q^d$ as input and
 outputs $F{\bf x}$.
 Our first contribution  is a generic construction
  of MSVC schemes for
outsourcing the computation   of $F{\bf x}$.
In order to protect $F$ and $\bf x$ from the servers in a $k$-server scheme, we choose
$a$ matrices $F_1, \ldots,F_a$ uniformly at random and  subject to
$F_1+ \cdots +F_a=F$;
choose
$b$ vectors ${\bf x}_1, \ldots,{\bf x}_b$ uniformly at random and  subject to
${\bf x}_1+ \cdots +{\bf x}_b={\bf x}$;  and  distribute the chosen matrices and
vectors among all servers such that: (1) each server only learns a proper subset of
the matrices; (2) each server only learns a proper subset of the vectors;
(3) the servers can perform computations on their respective shares in order to generate a set of results, from which the client is able to extract $F{\bf x}$.
While (1) and (2) allow us to attain the function privacy and input privacy easily, the method of distributing function/input shares should be properly designed such that (3)
is also satisfied.
Suppose that for every $\ell\in[k]$, the $\ell$-th server is given
$\rho_\ell=\{F_u: u\in A_\ell\}$ and  $\sigma_\ell=\{{\bf x}_v: v\in B_\ell\}$,
where $A_\ell\subsetneq [a]$ and $B_\ell\subsetneq [b]$.
Then  (3) will be satisfied if   there exist subsets
$C_1 \subseteq A_1\times B_1,\ldots, C_k\subseteq A_k\times B_k$ such that $\cup_{\ell=1}^k C_\ell=[a]\times [b]$.
In fact, when the equality holds, the $\ell$-th server only needs to compute and return    $y_\ell=\{F_u{\bf x}_v: (u,v)\in C_\ell\}$ and the client will be able to
reconstruct $F{\bf x}$ as $F{\bf x}=\sum_{\ell=1}^k \sum_{(u,v)\in C_\ell}F_u{\bf x}_v$.
In order to make the computation of each $F_u{\bf x}_v$
verifiable, we need a   {\em critical observation} on
matrix-vector multiplication \cite{BLR90,BW94}:
 if we choose a vector
${\bf  r}\leftarrow \mathbb{Z}_q^m$
 uniformly   at random   and  set
 ${\bf  s}_u= {\bf  r}  F_u$, then
the   output ${\bf  y}_{u,v}=F_u{\bf x}_v$ will satisfy the equality
$
  {\bf  s}_u \cdot {\bf  x}_v =  {\bf  r} \cdot {\bf  y}_{u,v},
 $ (${\bf  s}_u\cdot {\bf  x}_v$ stands for the inner  product of  ${\bf s}_u$ and
  ${\bf x}_v$). Without knowing $({\bf r}, {\bf s}_{u})$, an adversary will not be able to choose  a vector
  $\hat{\bf y}_{u,v}\neq {\bf y}_{u,v}$  such that
  $
  {\bf  s}_u \cdot {\bf  x}_v =  {\bf  r} \cdot \hat{\bf  y}_{u,v},
 $ except with probability $1/q$.
Furthermore, the verification takes $O(m+d)$ arithmetic operations modulo $q$,
which are significantly faster than the $O(md)$ arithmetic  operations required by the
naive computation of $F_u {\bf x}_v$.
 By properly applying  this verification  technique to all computations, we
obtain the   generic construction  with the expected security, input privacy,   function privacy, and practical
efficiency.

We optimize the generic construction in two directions: (i) minimizing the number
$k$ of required servers; (ii) minimizing the workload of the client and the server.
Minimizing $k$ is meaningful as that will make the  assumption of
 non-colluding servers most  practical. In order to achieve all
 expected properties of security and input/function privacy,
we show that
 the smallest $k$
is 3  and then give $\Pi_{\rm s}$, an instantiation of the generic construction with 3 servers.
In our generic construction, the client's workload is
dominated by $O(ab(m+d))$ arithmetic operations modulo $q$; the servers'
total workload is dominated by $O(abmd)$ arithmetic operations modulo $q$.
By appropriately choosing the parameters
we show that when $k=4$, $ab$ can be minimized. We then give $\Pi_{\rm w}$,  an instantiation of
the generic construction  with 4 servers.

\vspace{2mm}
\noindent
{\bf  Polynomial outsourcing.}
 In the literature of SSVC,  achieving  input  privacy
 has been highly non-trivial, especially when the   function $f$ requires
high-degree computations on the input $x$.
On one hand, in order to keep
$x$ private, the client somehow has to encrypt $x$
as ${\sf Enc}(x)$ with a semantically secure encryption   $\sf Enc$.
On the other hand, the server has to compute
$f$ on   ${\sf Enc}(x)$, but without decrypting ${\sf Enc}(x)$.
The latter situation requires that $\sf Enc$ should be
  homomorphic. As such an $\sf Enc$ would require heavy public-key operations,
  the SSVC schemes are hardly  practical.
In this paper, we try to tackle this problem with our generic construction.
 We observe that the evaluation of  many polynomials, including
univariate polynomials, bivariate polynomials,
 quadratic multivariate polynomials, and multivariate  polynomials  that have bounded degree in each variable,
  can be decomposed into
   a two-stage computation, where the first stage is a heavy matrix-vector multiplication and the second stage is a light inner product computation.
   By delegating the first-stage computation  with our MSVC schemes (say   $\Pi_{\rm s}$ and $ \Pi_{\rm w}$) for matrices and
   performing  the  second-stage computation
on its own, the client is able to
  offload most workload to  the cloud and   achieve all of the expected properties of
information-theoretic security and input/function privacy, and practical efficiency.

\vspace{2mm}
\noindent
{\bf Performance.}
We   implemented both  $\Pi_{\rm s}$ and   $\Pi_{\rm w}$ on a DELL Precision Tower T7810 that runs with  an  Intel  Xeon   E5-2650 Processor (2.30 GHz) and a RAM of 128GB.
We set
 $q$ to be a 256-bit   prime and consider the multiplication of a random matrix $F$ with a random vector $\bf x$.
For
$m=d=3000$, our experiments show that the client-side computations in $\Pi_{\rm s}$
and $\Pi_{\rm w}$ require around 177ms and 90ms, respectively. On the other hand, the naive computation of $F{\bf x}$ require around 2600ms.
The client-side computations in our MSVC schemes are significantly faster than the naive computation. Our MSVC schemes outperform
many existing SSVC schemes, which have been considered as practical.
For example, for $m=d=100$, the client in Pinocchio  \cite{PHGR13}
requires 10ms, which is
worse than the  5.73ms and 3.31ms used by $\Pi_{\rm s}$ and $\Pi_{\rm w}$,
 respectively. These experiments  show that our schemes are among the
most practical   schemes for  outsourcing computation.

\subsection{Related Work}

The study of efficient verification of arbitrary computations
 dates back to the work on interactive proofs \cite{Bab85,GMR85},
 probabilistically checkable proofs (PCPs)
 \cite{Kil92,Kil95}, computationally sound (CS) proofs \cite{Mic94}  and
the muggle proofs \cite{GKR08,Tha13}. These constructions yield {\em single-server}
schemes for securely outsourcing computations, which  are either   {\em interactive} or
 require    {\em random oracles}, and achieve {\em no} input/function privacy.

The {\em single-server} schemes that are both {\em non-interactive} and {\em free of random oracles} have been extensively
studied in the past decade.  Among them is  the single-server verifiable
computation (SSVC) of \cite{GGP10,AIK10,BF12,CKV11}, which enables a client to
delegate the computation of any boolean circuit $f$.
These  schemes
keep   the input $x$ semantically secure from
an  untrusted cloud server but heavily depend on
FHE and GCs.
They are rather   impractical.
There is  a long list of SSVC  schemes
  \cite{BGV11,FG13,CKLR11,EOAM16,FG12,GKR08,PST13,PRV12}  that focus on the delegation of
  specific functions such as polynomials and matrices.  These schemes are free of FHE and GCs but still require the server/client to perform heavy public-key operations such as modular exponentiations and paring computations, which are not as practical as the arithmetic operations in our schemes.  Furthermore, they don't  achieve any input privacy or function privacy.
In the SSVC schemes of \cite{FGP14,JY14,LPJY13},   the input $x$ is encrypted and  stored  on
cloud servers  and the function $f$ is provided  by the client.
These schemes   only achieve one-side privacy, such as the
input privacy, or function privacy, but not both.
The client and the server in these schemes must perform
a large number of expensive public-key operations as well.

Verifiable computation schemes in different {\em multi-server} models such as
 \cite{ACG+14,CRR11} have been studied as well.
In the model of Ananth et. al. \cite{ACG+14},
the client and all servers form a directed cycle and each
player has to receive a message from its predecessor and send
 a message to its successor, i.e., the communications between servers are required.
The computationally secure schemes of \cite{ACG+14} are input private and  FHE-free. However,
they still heavily depend on GCs and so are not as practical as the schemes of this paper.
In the model of Canetti et al.  \cite{CRR11}, the servers do not need to
communicate with each other and the client is a referee that
determines which server is honest.
The schemes of \cite{CRR11} are neither input private nor function private.
The security of \cite{ACG+14,CRR11} requires that
all servers do not collude with each other and at least one of the servers is honest.
In contrast, when considering security, we neither need non-colluding servers nor require
at least one honest server.
``Non-colluding" is not needed in our MSVC unless the input/function privacy is considered.
Ben-Or et al. proposed an MIP model  \cite{BGKW88}, where
the security  is guaranteed only as long as no two servers collude
and the malicious servers can potentially prevent honest ones from convincing the
client. In that model, and assuming collisions resistant hash functions, Canetti et al.
\cite{CRR112}
constructed a computationally secure delegation protocol whose number of rounds is
logarithmic in the time needed to compute $f(x)$.

\subsection{Organization}

Our MSVC model  is defined in Section 2; in Section 3,
we  propose the generic construction of MSVC  schemes for matrix outsourcing;
 we also obtain a scheme with the smallest number of servers and
a scheme  with  least workload;
we implement both schemes and give  performance analysis;
in Section 4 we    obtain  the new schemes for
polynomial outsourcing and
 outsourced multi-server PIR schemes where the servers' work is verifiable.
 Section 6 contains our concluding remarks.

\section{ Model and Definitions}

For a set $S$, we denote with ``$x\leftarrow S$" the process of choosing $x$ uniformly
at random from $S$. For a probabilistic algorithm $\cal A$, we denote with
``$ x\leftarrow  {\cal A}(\cdot)$" the process of running $\cal A$ on some appropriate
input  and assigning its output to $x$.
For an integer $n>0$, we denote with ``$[n]$" the set $\{1,\ldots,n\}$.
We use uppercase letters for matrices (e.g. $F$),
and lowercase bold letters for vectors (e.g. $\bf  x$).
We denote with $F^\top$ the transpose of a matrix $F$.
We  denote with $F[i,j]$ the $(i,j)$-entry of
the matrix $F$ and denote with ${\bf x}[i]$ the $i$-th entry of
$\bf x$.
Given two vectors ${\bf u}=(u_1,\ldots,u_m), {\bf v}=(v_1,\ldots,v_m)$, we denote with
${\bf u}\cdot {\bf v}$ the dot product of $\bf u,v$, i.e., ${\bf u}\cdot {\bf v}=\sum_{i=1}^m u_iv_i$.

In this paper, we work in a new multi-server verifiable computation (MSVC) model. Informally,
a  $k$-server verifiable computation scheme   is a   protocol between a client and $k$ servers. The client provides both $k$ shares
  of
the function $F$ and  $k$ shares   of the input
$x$   to all servers. Each server is expected to perform a set of  computations on
its shares, and produce an output, such that the $k$ outputs   together enable the client to reconstruct
  $F(x)$. Furthermore, the servers' results can be verified in order to guarantee correct reconstruction.
 The goal of MSVC is to make the client's
 work  as efficient as possible, and in particular much faster than
  the naive computation of    $F(x)$.
Let $\cal F$ be a function family.
A {\em $k$-server verifiable computation}   scheme $\Pi=({\sf  KeyGen, }$ ${\sf ProbGen,}$ ${\sf Compute,}$ ${\sf Verify})$
 for   $\cal F$  consists of the following
algorithms:
\begin{itemize}
\item ${\sf  KeyGen}(\lambda, F)$:
This is a randomized {\em key generation}  algorithm. It takes
a security parameter $\lambda$  and a function $F$ as input and   produces
a    value $PK_F$, which  will be used by a client to prepare its input,
$k$ function shares  $\rho_1,\ldots,\rho_k$, which  will be used by the servers to perform their respective computations,
 a value  $VK_F$, which will be used by the client to perform verification
 and reconstruct the function's output.

\item $ {\sf  ProbGen}(PK_F,x)$:
  This is a    {\em problem generation} algorithm. It takes $PK_F$
  and $x\in {\rm Domain}(F)$ as input and produces  $k$  input  shares  $\sigma_1,\ldots,\sigma_k$,
which will be given to the servers to compute with,  and a
 value $VK_x$,  which will be used for  verification.

\item ${\sf  Compute}(\ell, \rho_\ell, \sigma_\ell)$: This is the
{\em server-side}  algorithm. For every $\ell\in[k]$, it performs
    a set of computations on the function share
    $\rho_\ell$ and the  input share  $x_\ell$, and produces
    an output $y_\ell$.

\item  ${\sf  Verify}(VK_F, VK_{x},
\{y_\ell\}_{\ell=1}^k)$:
  This is the {\em verification} algorithm.
  It uses the  keys $VK_F, VK_x$  to   determine if
 $\{y_\ell\}_{\ell=1}^k$ form  a valid encoding of $F(x)$.
 If  $\{y_\ell\}_{\ell=1}^k$ is valid,  this algorithm
  converts  the servers' results
to  $y= F(x)$  and outputs $y$; otherwise, this algorithm outputs
$\perp$ (indicating that some servers are cheating).

 \end{itemize}

Our MSVC model   will consist of two phases: a {\em preprocessing} phase and
a {\em computing} phase. In the preprocessing phase,
 the client takes a function $F$ as input, runs the
 key generation algorithm  to produce the values  $PK_F,~ \{\rho_\ell\}_{\ell=1}^k,~ VK_F$, and   then sends $\rho_\ell$ to the $\ell$-th   server.
 In the computing phase,
 the client takes $PK_F$ and a value $x\in {\rm Domain}(F)$ as input,  runs
the problem generation algorithm to produce the values $\{\sigma_\ell\}_{\ell=1}^k, VK_x,$ and then sends   $\sigma_\ell$ to the $\ell$-th   server.
The $\ell$-th  server runs the server-side algorithm
to produce an output $y_\ell$.
Then the   client  uses   $VK_F, VK_x$ to verify   if  $\{y_\ell\}_{\ell=1}^k$ is a valid encoding of
$F(x)$, and depending on the verification result either outputs
$F(x)$ (when valid) or $\perp$ (when invalid).

%
%
%
%
%
%

The   scheme $\Pi$ is said to be {\em publicly delegatable} if
  $PK_F$ is public such that any client,  without executing
$\sf KeyGen$,  is able to run $\sf ProbGen$ to prepare its input.
Otherwise, the scheme is said to be privately delegatable.
The   scheme $\Pi$ is said to be {\em publicly verifiable} if
$VK_F$ and $VK_x$ are public such that
any client, without executing
$\sf KeyGen$ or $\sf ProbGen$, is able to run $\sf Verify$ to
determine whether $\{y_\ell\}_{\ell=1}^k$ is a valid encoding of $F(x)$.
Otherwise, the scheme is said to be privately verifiable.
In this paper, we shall construct publicly delegatable and privately verifiable
schemes.

The    scheme $\Pi$   is said to be   correct if
 $\sf KeyGen$ and $\sf ProbGen$  produce  values that always enable
the honest servers to compute values that will
verify successfully and be converted into the correct value of $F(x)$.
\begin{definition}
\label{def:correct}
{\bf (Correctness)}
The    scheme $\Pi$ is  {\em correct} if for any
   $F\in {\cal F},$    any   $
(PK_F, \{\rho_\ell\}_{\ell=1}^k, VK_F)\leftarrow {\sf KeyGen}(\lambda, F)$,
  any   $ x\in {\rm Domain}(F),$
  any  $ (\{\sigma_\ell\}_{\ell=1}^k,VK_x) \\ \leftarrow {\sf ProbGen}(PK_F,x)$
 and   any $\{y_\ell\leftarrow {\sf Compute}(\ell,\rho_\ell,$ $\sigma_\ell)\}_{\ell=1}^k$,
it holds that \\
${\sf Verify}(VK_F, VK_x,\{y_\ell\}_{\ell=1}^k)=F(x)$.
\end{definition}

\begin{figure}[h]
{
\begin{center}
\begin{boxedminipage}{12cm}
\begin{itemize}
\item[] \hspace{-1cm}    $(PK_F, \{\rho_\ell\}_{\ell=1}^k, VK_F)\leftarrow
{\sf KeyGen}(\lambda, F)$;
\item[] \hspace{-1cm} for $h=1$ to $p$ do (remark: $p$ is   the number of attempts  that $\cal A$ can make)
\begin{itemize}
\item[] \hspace{-1cm}  $x^{(h)}\leftarrow {\cal A}\big(PK_F, \{\rho_\ell\}_{\ell=1}^k,
\big\{\{\sigma^{(t)}_{\ell}\}_{\ell=1}^k, \{\hat{y}^{(t)}_{\ell}\}_{\ell=1}^k, b_t\big\}_{t=1}^{h-1}\big)$;
\item[] \hspace{-1cm}  $(\{\sigma^{(h)}_{\ell}\}_{\ell=1}^k, VK_{x^{(h)}})\leftarrow {\sf ProbGen}
(PK_F, {x}^{(h)})$;
\item[] \hspace{-1cm}  $\{\hat{y}^{(h)}_{ \ell}\}_{\ell=1}^k\leftarrow
 {\cal A}\big(PK_F, \{\rho_\ell\}_{\ell=1}^k,  \big\{\{\sigma^{(t)}_{ \ell}\}_{\ell=1}^k,
 \{\hat{y}^{(t)}_{ \ell}\}_{\ell=1}^k, b_t\big\}_{t=1}^{h-1},\{\sigma^{(h)}_{ \ell}\}_{\ell=1}^k\big)$;
\item[] \hspace{-1cm}  $\hat{y}^{(h)}\leftarrow {\sf Verify}\big(VK_F, VK_{x^{(h)}},
\{\hat{y}^{(h)}_{ \ell}\}_{\ell=1}^k\big)$;
\item[] \hspace{-1cm} if $\hat{y}^{(h)}=\perp$, set $b_h=0$; otherwise, set
$b_h=1$;
\end{itemize}
\item[] \hspace{-1cm}   if there is an $h\in [p]$
such that   $\hat{y}^{(h)}\notin \{F(x^{(h)}),\perp\}$, then   output 1;
  otherwise,  output 0.
\end{itemize}
\end{boxedminipage}

\vspace{2mm}

  Fig. 1: Experiment ${\sf Exp}_{{\cal A}, \Pi}^{{\rm verif}}( F, p)$

\vspace{-4mm}
\end{center}
}
\end{figure}
In our MSVC model, the scheme $\Pi$ is considered as   {\em secure} if no malicious
servers can  persuade the verification
algorithm to output  a   result $\hat{y}\notin \{F(x),\perp\}$. This intuition can be formalized
with an   experiment ${\sf Exp}_{{\cal A}, \Pi}^{{\rm verif}}(F,p)$.
 In this experiment (Fig. 1),
 the challenger firstly runs the key generation algorithm to produce the
 necessary keys $PK_F, \{\rho_\ell\}_{\ell=1}^k, VK_F$  to initialize the scheme.
 The adversary $\cal A$ then makes  $p$ attempts  to
 choose a function input, learn the  encoding of that input, craft the servers'
 results for that input, and then see if these crafted results will
  be able to  cause the  verification algorithm to output a wrong function value.
 The adversary succeeds   if its crafted results ever cause $\sf Verify$
 to output a wrong value.
For the scheme  $\Pi$ to be secure,
$\cal A$ is  allowed to succeed only   with a very small probability $\epsilon$.

 \begin{definition}
\label{def:secure}
{\bf (Security)}
The   scheme $\Pi$ is  $(p,\epsilon)$-{\em secure} if for all    $F\in {\cal F}$,
for any adversary $\cal A$, it holds that
$
\Pr\left[{\sf Exp}_{{\cal A}, \Pi}^{{\rm verif}}( F,p)=1\right]\leq \epsilon,
$
where the probability is taken over the randomness used by $\cal A$ and   the experiment.
\end{definition}

In our definition of security we do not limit the computational power of the adversary
$\cal A$. Consequently, our security will be information-theoretic.
However, we do upper bound the number of attempts that can be made by
$\cal A$. In particular, the success probability $\epsilon$
will be  bounded by a function of $p$.
In order to compare with the computationally secure SSVC schemes, we usually
require that $\epsilon$ should be negligible in the statistical security parameter
$\lambda$, as long as $p$ is a polynomial function of $\lambda$.

Intuitively, the     scheme $\Pi$ is said to be  input private if
each individual server learns absolutely no information about
the client's input. This property will be captured by the requirement
that each individual server should receive an input share that is statistically independent of
the client's input.
 \begin{definition} \label{def:ipri}
{\bf (Input privacy)}
The    scheme $\Pi$ is  {\em input private} if
for any
   $F\in {\cal F},$
any $x^{(0)}, x^{(1)}\in {\rm Domain}(F)$,  any
  $\ell^*\in[k]$, any
  $\{(\{\sigma^{(b)}_\ell\}_{\ell=1}^k, VK_{x^{(b)}})\leftarrow {\sf ProbGen}(PK_F,
x^{(b)})\}_{b=0}^1$,
$\sigma_{\ell^*}^{(0)}$
and $\sigma_{\ell^*}^{(1)}$ are  identically distributed.
\end{definition}

Intuitively, the  scheme $\Pi$ is said to be  function private  if
each individual server learns absolutely no information about
the client's function. This property will be captured by
the requirement that each individual server should receive a function share
that is statistically independent of the client's function.
\begin{definition} \label{def:fpri}
{\bf (Function privacy)}
The    scheme $\Pi$ is  {\em function private} if
  for any $F^{0}, F^{1}
\in {\cal F}$,  for any $\ell^*\in [k]$,
and any
  $\{(PK_{F^{(b)}}, \{\rho^{(b)}_\ell\}_{\ell=1}^k, VK_{F^{(b)}})\leftarrow {\sf KeyGen}(\lambda, F^{(b)})\}_{b=0}^1$,
$(PK_{F^{(0)}}, \rho^{(0)}_{\ell^*})$
and $(PK_{F^{(1)}}, \rho^{(1)}_{\ell^*})$ are identically distributed.

\end{definition}

The  scheme  $\Pi$ is said to be  outsourceable  if the client's work in the computing phase is substantially faster than  the naive computation of
the function.
\begin{definition}
\label{def:efficient}
{\bf (Outsourceable)}
The   scheme $\Pi$ is  {\rm outsourceable} if it permits efficient problem generation  and result verification.
That is, for any $F\in {\cal F}$ and any   $x\in {\rm Domain}(F)$,  the total time
 $T_{\rm c}$ required for
 ${\sf ProbGen}(PK_F,{x})$ and ${\sf Verify}(VK_F, VK_x, \{y_\ell\}_{\ell=1}^k)$ is
  $o(T_{\rm n})$, where $T_{\rm n}$ is the time required by the naive computation of
 $F(x)$.
\end{definition}

As in the existing SSVC   protocols  \cite{BGV11,GGP10}, the client's
work in the preprocessing phase may be as expensive as the naive computation of
the function.
However, executing $\sf KeyGen$ is a one-time computation that can be amortized over
the computation  of $F$ on many different inputs, and thus acceptable.
In other words, we will also work in the   {\em amortized} model of  \cite{BGV11,GGP10}.

\section{Multi-Server  Schemes for Matrix Outsourcing}
\label{sec:con}

In this section, we propose   an  MSVC scheme   for
outsourcing the matrix-vector multiplications of the form $F{\bf x}$.
The scheme provides public delegation and  private verification; it is
information-theoretically secure,  input
 private and function private.
The function family supported by  our scheme is ${\cal F}=\mathbb{Z}_q^{m\times d}$,  the set of all $m\times d$ matrices
  over a finite field $\mathbb{Z}_q$. Each matrix $F\in {\cal F}$ is interpreted as a function
that   takes a    vector
${\bf  x} \in \mathbb{Z}_q^d$ as input and outputs  ${\bf y}=
F {\bf  x}\in \mathbb{Z}_q^m$.
In a nutshell, the efficient verification in our scheme is
based on the following technical lemma:
\begin{lemma}
\label{lem:vm}
Let $\hat{\bf  y}, {\bf  y} \in \mathbb{Z}_q^{m}$ be distinct  vectors. Then
$\Pr_{{\bf  r}\leftarrow \mathbb{Z}_q^m}\left[   {\bf  r} \cdot \hat{\bf  y}
= {\bf  r} \cdot {\bf  y}   \right]\leq \frac{1}{q}. $
\end{lemma}
To the best of our knowledge, Lemma \ref{lem:vm}  dates back to \cite{BLR90,BW94}
 and has been
used   to construct SSVC schemes \cite{Moh11}.
Here are the {\em critical observations}:
\begin{itemize}
\item   If one chooses
${\bf  r}\leftarrow \mathbb{Z}_q^m$ and keeps both $\bf  r$ and ${\bf  s}=
{\bf  r}  F$ secret, then
for any ${\bf  x}\in \mathbb{Z}_q^d$,
the  function value ${\bf  y} =F{\bf x}$ will satisfy the following equation
due to the associative law of matrix multiplications:
\begin{equation}
\label{eqn:vf1}
  {\bf  s}\cdot {\bf  x} =  {\bf  r} \cdot {\bf  y}.
 \end{equation}

\item In a protocol where the server computes ${\bf y}=F{\bf x}$, the client can simply confirm the correctness of $\bf y$ after seeing that it satisfies
    (\ref{eqn:vf1}); and  a server responding with  $\hat{\bf  y}\neq F  {\bf  x}$ will pass  the verification of (\ref{eqn:vf1}) with
 probability  at most $1/q$, because $\hat{\bf y}$  verifies if and only if
$ {\bf  r} \cdot \hat{\bf  y}
= {\bf  r} \cdot {\bf  y}$.
\item  The  verification of (\ref{eqn:vf1})   can be done with
 $O(m+d)$ arithmetic operations; it is significantly faster than the naive computation of   $F {\bf  x}$, which requires  $O(md)$ arithmetic  operations.
   \end{itemize}
 These  observations give  us an inspiring SSVC  scheme: in the preprocessing phase, the client
sends the matrix $F$ to the  server and keeps  a  private key   $({\bf r,s})$ for
verification; in the computing phase, client simply sends
$\bf x$ to the   server; the server returns ${\bf y}=F{\bf x}$;
and finally the client checks
 (\ref{eqn:vf1}).
 However, this scheme is neither input private nor function private.

 \subsection{Generic Construction}

We try to achieve both the input privacy and function privacy in
 the MSVC model. The basic idea is secret-sharing both  the function
$F$ and   the input $\bf x$ among   multiple servers such that each  server
learns absolutely no information about $F$ or $\bf x$,  but the servers are still able to perform
certain  computations on their shares and the computation results  together enable the reconstruction of  $F{\bf x}.$ We shall propose a generic construction
and then instantiate it with various   parameters to  attain  the best efficiency.

In our generic construction, it suffices to use an {\em additive secret sharing} \cite{Bei96}
 where the  secret $\alpha$ is an element of   a ring
 $R$, the shares $\alpha_1,\ldots,\alpha_n\leftarrow R$ are uniformly chosen subject to
  $\alpha_1+\cdots+\alpha_n=\alpha$, and each server is given a proper subset of
 the shares $\{\alpha_1,\ldots,\alpha_n\}$
 such that any  single server learns absolutely no information about $\alpha$ and
  a subset of the servers together are able to reconstruct $\alpha$
 if and only if their shares cover all of
 $\alpha_1,\ldots,\alpha_n$.

Let $a,b,k>1$ be integers and let $A=[a]$ and $B=[b]$. In a $k$-server scheme,  we  shall  decompose the function
  as $F=F_1+\cdots+F_a$, decompose the function input    as $ {\bf x}={\bf x}_1+\cdots +{\bf x}_b$,
and distribute the additive shares $\{F_u:u\in A\}$ and $\{{\bf x}_v: v\in B\}$ among  $k$ servers ${\cal S}_1, \ldots, {\cal S}_k$.
For every $\ell\in [k]$, let
\begin{equation}
\label{eqn:al}
\begin{split}
A_\ell&=\{u\in A: F_u {\rm~is~given~to~}{\cal S}_\ell\};  \\
B_\ell&=\{v\in B: {\bf x}_v {\rm~is~given~to~}{\cal S}_\ell\}.
\end{split}
\end{equation}
Then each server ${\cal S}_\ell$ is able to compute
$F_u {\bf x}_v$ for all $(u,v)\in A_\ell\times B_\ell$.
Because
\begin{equation}
F{\bf x}=\sum_{u=1}^a F_u \cdot \sum_{v=1}^b {\bf x}_v=\sum_{(u,v)\in A\times B} F_u{\bf x}_v,
\end{equation}
the $k$ servers'   results, i.e., $\{F_u{\bf x}_v: (u,v)\in \cup_{\ell=1}^k  (A_\ell\times B_\ell)\}$,
suffice to reconstruct $F{\bf x}$  if and only if
\begin{equation}
\label{eqn:cov}
(A\times B) \subseteq \bigcup_{\ell=1}^k (A_\ell\times B_\ell),
\end{equation}
i.e., $\{A_\ell\times B_\ell\}_{\ell=1}^k$ form a cover of  $A\times B$.
For every $\ell\in[k]$,
the function $F$ is secret  from
${\cal S}_\ell$ if and only if
\begin{equation}
\label{eqn:A}
A_\ell\neq A.
\end{equation}
For every $\ell\in [k]$,  the input $\bf x$ is secret from
${\cal S}_\ell$ if and only if
\begin{equation}
\label{eqn:B}
B_\ell \neq B.
\end{equation}
When  $\{A_\ell\times B_\ell\}_{\ell=1}^k$ form a cover of  $A\times B$,
there must exist $k$ sets  $C_1,\ldots, C_k$ such that
\begin{equation}
\label{eqn:par}
\begin{split}
&C_\ell \subseteq A_\ell\times B_\ell {\rm~for~every~}\ell\in [k];\\
&C_\ell\cap C_{\ell^\prime}=\emptyset {\rm~for~all~}\ell\neq \ell^\prime; {\rm~and}\\
&C_1\cup\cdots \cup C_k=A\times B,
 \end{split}
 \end{equation}
  i.e.,
$\{C_1,\ldots,C_k\}$ form a partition of $A\times B$.
Based on (\ref{eqn:cov}), (\ref{eqn:A}), (\ref{eqn:B}) and (\ref{eqn:par}), our generic construction works as follows:
in the preprocessing phase, the client sends $\{F_u: u\in A_\ell\}$ to  ${\cal S}_\ell$ for every
$\ell\in [k]$; in the computing phase, the client sends $\{{\bf x}_\ell: \ell\in B_\ell\}$ to ${\cal S}_\ell$
for every $\ell\in [k]$;  the server ${\cal S}_\ell$ returns
 $\{F_u{\bf x}_v: (u,v)\in C_\ell\}$.
The verification will be done with   Lemma \ref{lem:vm}.

Let $k,a,b,A,B,\{A_\ell\}_{\ell=1}^k,\{B_\ell\}_{\ell=1}^k$ and
 $\{C_\ell\}_{\ell=1}^k$ be    the parameters and sets  that satisfy   (\ref{eqn:cov}), (\ref{eqn:A}), (\ref{eqn:B}) and (\ref{eqn:par}).
Our generic construction
$\Pi=({\sf KeyGen, ProbGen, }$ ${\sf Compute, Verify})$ of
 a  $k$-server verifiable computation scheme
for matrix outsourcing    can be specified  as below.
\begin{itemize}
\item $  {\sf KeyGen}(\lambda,F)$: This algorithm takes the security parameter $\lambda$ and a matrix $F\in \mathbb{Z}_q^{m\times d}$ as input.
    It chooses $F_1, \cdots, F_a\leftarrow \mathbb{Z}_q^{m\times d}$ uniformly   subject to
     $F_1+\cdots +F_a=F$;         chooses
    ${\bf r}\leftarrow \mathbb{Z}_q^m$, computes   ${\bf s}_u={\bf r}  F_u$
    for every $u\in A$, defines $\rho_\ell=\{F_u: u\in A_\ell\}$
    for every $\ell\in[k]$,  and finally outputs
$PK_F=\perp, \{\rho_\ell\}_{\ell=1}^k, $
and  $ VK_F=({\bf r},
  \{{\bf s}_u\}_{u=1}^a). $

 \item  $ {\sf ProbGen}(PK_F,{\bf x})$: This algorithm takes $PK_F=\perp$ and
 a vector ${\bf  x} \in \mathbb{Z}_q^d$ as input. It chooses
     ${\bf x}_1,\ldots,{\bf x}_b\leftarrow \mathbb{Z}_q^d$ uniformly  subject to   ${\bf x}_1+\cdots +
     {\bf x}_b={\bf x}$,  defines $\sigma_\ell=\{{\bf x}_v: v\in B_\ell\}$
     for every $\ell\in[k]$, and  finally    outputs
     $\{\sigma_\ell\}_{\ell=1}^k,$
     and $  VK_{\bf  x}=({\bf  x}_1,\ldots,
     {\bf  x}_b).$

\item $ {\sf Compute}(\ell,\rho_\ell,\sigma_\ell)$: For every $\ell\in [k]$,
this algorithm takes   a set  $\rho_\ell=\{F_u: u\in A_\ell\}$ of function shares and   a set
$\sigma_\ell=\{{\bf x}_v: v\in B_\ell\}$  of input shares   as input.
It computes ${\bf y}_{u,v}=F_u {\bf x}_v$ for all $(u,v)\in C_\ell$ and outputs
$y_\ell=\{{\bf y}_{u,v}: (u,v)\in C_\ell\}.$

\item ${\sf Verify}( VK_F, VK_{\bf x}, \{y_\ell\}_{\ell=1}^k)$:
This algorithm takes  $VK_F=({\bf r},\{{\bf s}_u\}_{u=1}^a), \\ VK_{\bf x}=({\bf x}_1,\ldots,{\bf x}_b)$
and the $k$ servers' computation results $\{y_\ell\}_{\ell=1}^k$ as input.
For every $\ell\in [k]$ and $(u,v)\in C_\ell$, it checks the   equality
\begin{equation}
\label{eqn:verg}
{\bf r}\cdot {\bf y}_{u,v}={\bf s}_u\cdot {\bf x}_v.
\end{equation}
If (\ref{eqn:verg}) always holds,
the algorithm  outputs
${\bf y}=
\sum_{\ell=1}^k \sum_{(u,v)\in C_\ell} {\bf y}_{u,v};
$
otherwise, it  outputs $\perp$.

\end{itemize}

\vspace{0mm}
\noindent
{\bf Correctness.}
The correctness of  $\Pi$ requires that for
any function $F\in \mathbb{Z}_q^{m\times d}$, any   input    ${\bf x}
\in \mathbb{Z}_q^d$, any  $(PK_F,\{\rho_\ell\}_{\ell=1}^k, VK_F)\leftarrow {\sf KeyGen}(\lambda,F)$,
and any $(\{\sigma_\ell\}_{\ell=1}^k, \\ VK_{\bf x})\leftarrow {\sf ProbGen}(PK_F,{\bf x})$,
if $y_\ell $ is output by ${\sf Compute}(\ell,\rho_\ell,\sigma_\ell)$ for all $\ell\in[k]$, then
${\sf Verify}(VK_F,VK_{\bf x}, \{\pi_\ell\}_{\ell=1}^k)$ will always output $F{\bf x}$.
When the server-side algorithm  ${\sf Compute}(\ell,\rho_\ell,\sigma_\ell)$  is honestly executed for all
$\ell\in[k]$,
for   every $(u,v)\in C_\ell$ and  $i\in[m]$, we must  have that
$
{\bf y}_{ u,v}[i]=\sum_{j=1}^d  F_{u}[i,j] \cdot {\bf x}_{v}[j].
$
It follows that
\begin{equation}
\begin{split}
{\bf r}\cdot {\bf y}_{ u,v}=\sum_{i=1}^m {\bf r}[i]\cdot {\bf y}_{ u,v}[i]
&=\sum_{i=1}^m {\bf r}[i] \sum_{j=1}^d  F_{u}[i,j]  \cdot {\bf x}_{v}[j]\\
&=
\sum_{j=1}^d {\bf x}_{v}[j] \sum_{i=1}^m {\bf r}[i] \cdot F_{u}[i,j] =
{\bf x}_v\cdot {\bf s}_u,
\end{split}
\end{equation}
which is exactly the equality (\ref{eqn:verg}). It follows that the verification algorithm will  output
\begin{equation*}
\sum_{\ell=1}^k \sum_{(u,v)\in C_\ell} {\bf y}_{u,v}=\sum_{(u,v)\in A\times B} F_u\cdot {\bf x}_v=
\sum_{u=1}^a F_u\cdot \sum_{v=1}^b {\bf x}_v=F{\bf x}.
\end{equation*}

\vspace{2mm}
\noindent
{\bf Input privacy.}
In the generic construction, each server ${\cal S}_\ell$ is given
a set $\sigma_\ell=\{{\bf x}_v: v\in B_\ell\}$ of input shares.
As $B_\ell$ is a proper subset of $B=[b]$ and all shares of $\bf x$ are chosen uniformly subject to
${\bf x}_1+\cdots+{\bf x}_b={\bf x}$,    $\sigma_\ell$ must be truly random and independent of $\bf x$.
 Any single server will learn absolutely no information about
 $\bf x$, even if it has unlimited computing power.
Hence, $\Pi$ achieves information-theoretic  input privacy (as defined in  Definition \ref{def:ipri}).

\vspace{2mm}
\noindent
{\bf Function privacy.}
In the generic construction, each server ${\cal S}_\ell$ is given
a set $\rho_\ell=\{F_u: u\in A_\ell\}$ of function shares.
As $A_\ell$ is a proper subset of $A=[a]$ and all shares of $F$ are chosen uniformly  subject to
$F_1+\cdots+F_a=F$, $\rho_\ell$ must be truly random and independent of
 $F$.
 Any single server will learn absolutely no information about
 $F$, even if it has unlimited computing power.
Hence, $\Pi$ achieves information-theoretic  function privacy (as defined in  Definition \ref{def:fpri}).

\vspace{2mm}
\noindent
{\bf Security.}
The security of an MSVC scheme requires that no   adversary   is able to persuade the
verification algorithm to both accept the dishonest servers' results   and
  output a wrong value, except with a very small probability.
In our generic construction the client requires each server to
compute  a set of matrix-vector multiplications
  and the verification of each matrix-vector multiplication is done with
the inspiring SSVC scheme. As a result, the security will follow from Lemma \ref{lem:vm}.

\begin{theorem}
\label{thm:sec}
The generic construction $\Pi$ is $(p, pab/(q-pab))$-secure. That is,
$$\Pr[{\sf Exp}_{{\cal A}, \Pi}^{\rm verif}(F,p)=1]
\leq \frac{pab}{q-pab}.$$
\end{theorem}

\begin{proof}
Let $F\in \mathbb{Z}_q^{m\times d}$ be any admissible function.
Let $\cal A$ be any   adversary that makes at most $p$
 attempts  in the security experiment. We show that
  $\Pr[{\sf Exp}_{{\cal A}, \Pi}^{\rm verif}(F,p)=1]\leq
pab/(q-pab)$. By Definition
\ref{def:secure}, the experiment ${\sf Exp}_{{\cal A}, \Pi}^{\rm verif}(F,p)$
 will be done between $\cal A$ and the challenger as follows:
\begin{itemize}
\item Given  $F$, the challenger  runs the key generation algorithm
${\sf KeyGen}(\lambda, F)$ and then invokes $\cal A$ as below:
\begin{itemize}
\item Choose $F_1, \cdots, F_a\leftarrow \mathbb{Z}_q^{m\times d}$ uniformly at random  subject  to
     $F_1+\cdots +F_a=F$;  choose
    ${\bf r}  \leftarrow \mathbb{Z}_q^m$ and set   ${\bf s}_u ={\bf r}  F_u$ for every
    $u\in A$; set $\rho_\ell=\{F_u: u\in A_\ell\}$ for every $\ell\in[k]$;
    \item  Invoke  $\cal A$ with  $PK_F=\perp$ and $ \{\rho_\ell\}_{\ell=1}^k$;
keep $VK_F=({\bf r},
  \{{\bf s}_u\}_{u=1}^a)$ secret.
\end{itemize}
\item  for $h=1$ to $p$ do (remark: $p$ is the total number of attempts that will be made by the adversary $\cal A$)
\begin{itemize}
\item   Based on the  current view $\big(PK_F, \{\rho_\ell\}_{\ell=1}^k,
\big\{\{\sigma^{(t)}_{\ell}\}_{\ell=1}^k, \{\hat{y}^{(t)}_{\ell}\}_{\ell=1}^k,
 b_t\big\}_{t=1}^{h-1}\big)$,  ${\cal A}$ chooses an input ${\bf x}^{(h)}\in \mathbb{Z}_q^d$ and gives it to the challenger;
\item  The challenger runs
${\sf ProbGen}(PK_F, {\bf x}^{(h)})$ as follows:
choose ${\bf x}^{(h)}_1,\ldots, {\bf x}^{(h)}_b \\ \leftarrow \mathbb{Z}_p^d$ uniformly at random
and subject to ${\bf x}^{(h)}_1+\cdots+{\bf x}^{(h)}_b={\bf x}^{(h)}$; define
$\sigma^{(h)}_\ell=\{{\bf x}^{(h)}_v: v\in B_\ell\}$ for every $\ell\in [k]$;
define $VK_{{\bf x}^{(h)}}=({\bf x}^{(h)}_1,\ldots, {\bf x}^{(h)}_b)$. Finally,
it gives   $\{\sigma^{(h)}_{\ell}\}_{\ell=1}^k$ to $\cal A$.
\item
Based on the current view   $\big(PK_F, \{\rho_\ell\}_{\ell=1}^k,
\big\{\{\sigma^{(t)}_{\ell}\}_{\ell=1}^k, \{\hat{y}^{(t)}_{\ell}\}_{\ell=1}^k,
 b_t\big\}_{t=1}^{h-1}$,  $\{\sigma^{(h)}_{\ell}\}_{\ell=1}^k\big)$,
 ${\cal A}$ crafts a set  $\{\hat{y}^{(h)}_{\ell}\}_{\ell=1}^k$ of server results and
gives them  to the challenger, where
$\hat{y}^{(h)}_{\ell}=\{\hat{\bf y}^{(h)}_{u,v}: (u,v)\in C_\ell\}$
for every $\ell\in [k]$;

\item   The challenger runs   ${\sf Verify}\big(VK_F, VK_{{\bf x}^{(h)}},
\{\hat{y}^{(h)}_{\ell}\}_{\ell=1}^k\big)$ to compute a value   $\hat{\bf y}^{(h)}$;
if  $\hat{\bf y}^{(h)}=\perp$, it sets $b_h=0$; otherwise, it sets $b_h=1$.
\end{itemize}
\item    If there is an $h\in [p]$ such that   $\hat{\bf y}^{(h)}\notin\{F{\bf x}^{(h)},\perp\}$, then   output 1;
  otherwise,  output 0.
\end{itemize}

 For every $h\in [p]$, let ${\bf E}_h$ be the event
 that $\hat{\bf y}^{(h)}\notin\{F{\bf x}^{(h)},\perp\}$.
 Then (i) ${\bf E}_h$ occurs if and only if $b_h=1$ and $\hat{y}^{(h)}\neq F{\bf x}^{(h)}$;
 and (ii)
 The event ${\sf Exp}_{{\cal A}, \Pi}^{\rm verif}(F,p)=1$ occurs if and only if
 $
\cup_{h=1}^p{\bf E}_h
$ occurs.
It suffices to show that $\Pr[\cup_{h=1}^p{\bf E}_h]\leq pab/(q-pab)$.

For every $h\in [p]$, the event $b_h=1$ occurs  if and only if
 for all $\ell\in [k]$ and $(u,v)\in C_\ell$, the equality
\begin{equation}
\label{eqn:verhat}
{\bf r}\cdot \hat{\bf y}^{(h)}_{u,v}={\bf s}_u \cdot {\bf x}_v^{(h)}
\end{equation}
is true.  On the other hand, for every
$h\in [p]$ and $\ell\in[k]$, let
$ y^{(h)}_{\ell} = \{{\bf y}^{(h)}_{u,v}: (u,v)\in C_\ell\} $
be the servers' results generated by executing   $\sf Compute$ faithfully. The correctness
of $\Pi$ implies that for all $\ell\in [k]$ and $ (u,v)\in C_\ell,$
\begin{equation}
\label{eqn:verproof}
{\bf r}\cdot {\bf y}^{(h)}_{u,v}={\bf s}_u \cdot {\bf x}_v^{(h)}
\end{equation}
must be true.
Due to  (\ref{eqn:verhat}) and (\ref{eqn:verproof}),   the event $b_h=1$ occurs  if and only if
$
{\bf r}\cdot \hat{\bf y}^{(h)}_{u,v}={\bf r}\cdot {\bf y}^{(h)}_{u,v}
$ is true
for all $\ell\in [k]$ and $(u,v)\in C_\ell$. Equivalently,  the event
$b_h=1$ occurs if and only if
$S_h:=\{\hat{\bf y}^{(h)}_{u,v}-{\bf y}^{(h)}_{u,v}: \ell\in [k], (u,v)\in C_\ell\}$ is
a set of solution vectors  of the following linear equation system
\begin{equation}
\label{eqn:sys}
{\bf r}\cdot {\bf y}=0,
\end{equation}
where ${\bf r}$ is the coefficient matrix and $\bf y$ is the vector of  unknowns.

Based on the specifications and the correctness  of $\Pi$,
for every $h\in [p]$, we have that
\begin{equation*}
\hat{\bf y}^{(h)}=\sum_{\ell=1}^k \sum_{(u,v)\in C_\ell} \hat{\bf y}^{(h)}_{u,v},
{\rm~and~}
\sum_{\ell=1}^k \sum_{(u,v)\in C_\ell} {\bf y}^{(h)}_{u,v}=F{\bf x}^{(h)}.
\end{equation*}
Then for every $h\in[p]$, the event $\hat{\bf y}^{(h)}\neq F{\bf x}^{(h)}$  occurs  only if
 there is at least one $\ell\in [k]$ and at least one $(u,v)\in C_\ell$ such that
$
\hat{\bf y}^{(h)}_{u,v}\neq {\bf y}^{(h)}_{u,v}.
$

 In order to  understand the events $b_h=1$ and
 $\hat{\bf y}^{(h)}\neq F{\bf x}^{(h)}$,
 for every $h\in [p]$ we define   three subsets
 \begin{equation*}
\begin{split}
X_h&=\{(u,v)\in A\times B: \hat{\bf y}^{(h)}_{u,v}-{\bf y}^{(h)}_{u,v}{\rm ~is~not~a~solution~of~} (\ref{eqn:sys}) \};\\
Y_h&=\{(u,v)\in A\times B: \hat{\bf y}^{(h)}_{u,v}-{\bf y}^{(h)}_{u,v}{\rm ~is~a~nonzero~solution~of~} (\ref{eqn:sys}) \};
\\
Z_h&=\{(u,v)\in A\times B: \hat{\bf y}^{(h)}_{u,v}-{\bf y}^{(h)}_{u,v}{\rm
~is~a~zero~solution~of~} (\ref{eqn:sys}) \},
\end{split}
\end{equation*}
which form a partition of  $A\times B$.
For every $h\in[p]$, the above analysis shows that:
(i) The event $b_h=1$ occurs  if and only if $X_h=\emptyset$;
(ii) The event $\hat{\bf y}^{(h)}\neq F{\bf x}^{(h)}$ occurs only if
$Y_h\neq \emptyset$.
 It follows that for every $h\in[p]$, the event
 ${\bf E}_h$ occurs only if  $X_h=\emptyset$ and
 $Y_h\neq \emptyset$.
For every $h\in [p]$, let ${\bf F}_h$ be the event that
 $h=\min\{t\in [p]: Y_t\neq \emptyset\}$. It is easy to see that
  $(\cup_{h=1}^p {\bf E}_h) \subseteq
(\cup_{h=1}^p  {\bf F}_h)$ and thus
\begin{equation}
\label{eqn:bound}
\Pr[\cup_{h=1}^p {\bf E}_h]\leq
\Pr[\cup_{h=1}^p {\bf F}_h]
=\sum_{h=1}^{p} \Pr[{\bf F}_h ].
\end{equation}
Note that ${\bf F}_h$ occurs only if none of the events ${\bf F}_1,\ldots,{\bf F}_{h-1}$
occurs, i.e., $Y_1= \cdots=Y_{h-1}=\emptyset$.
Hence, when  ${\bf F}_h$ occurs, we must have that $X_t\cup Z_t=A\times B$
for every $t\in [h-1]$.
For every $(u,v)\in Z_t$,
$\hat{\bf y}^{(h)}_{u,v}-{\bf y}^{(h)}_{u,v}$ is a zero solution of (\ref{eqn:sys}) and
gives $\cal A$ absolutely no information about $\bf r$;
for every  $(u,v)\in X_t$,
$\hat{\bf y}^{(h)}_{u,v}-{\bf y}^{(h)}_{u,v}$ is not a  solution of (\ref{eqn:sys}) and
gives  $\cal A$ exactly the information that
\begin{equation}
\label{eqn:testz}
{\bf r} \cdot (\hat{\bf y}^{(h)}_{u,v}-{\bf y}^{(h)}_{u,v})\neq 0,
\end{equation}
allows $\cal A$ to rule out $\leq q^{m-1}$ possibilities of $\bf r$.
When ${\bf F}_h$ occurs, $X_1\cup\cdots\cup X_{h-1}$  contains
$\leq (h-1)ab$ elements and helps $\cal A$ to rule out at most $(h-1)ab q^{m-1}$
possibilities of $\bf r$. As a result, in the $h$-th attempt,
the $\bf r$ is still uniformly distributed over the set of all remaining vectors that have not been ruled out.
By providing a set $\{\hat{\bf y}^{(h)}_{u,v}: \ell\in [k], (u,v)\in C_\ell\}$,
    $\cal A$ would define   $\leq ab$ new equation systems of the form
(\ref{eqn:testz})
that have a nonzero coefficient matrix.
The solution spaces of these nontrivial equation systems together  cover at most $abq^{m-1}$  vectors
out of the $q^m-(h-1)abq^{m-1}$ remaining vectors.
Given that none of the events ${\bf F}_1,\ldots,{\bf F}_{h-1}$ occurs, the secret vector  $\bf r$ is still uniformly distributed over the
set of all remaining vectors. Therefore,
we must have that
\begin{equation}
\label{eqn:bfh}
\Pr[{\bf F}_h]\leq \frac{abq^{m-1}}{q^m-(h-1)abq^{m-1}}\leq  \frac{ab}{q-p ab},
\end{equation}
 i.e., the  $\bf r$ will fall into the   union of these solution spaces
with probability at most $ab/(q-p ab)$.
Due to (\ref{eqn:bound}) and (\ref{eqn:bfh}), we have that
$
\Pr[{\sf Exp}_{{\cal A}, \Pi}^{\rm verif}(F,p)=1] \leq pab/(q-p ab).
$
\end{proof}

In Theorem \ref{thm:sec}, any adversary  making  $\leq p$ attempts succeeds in breaking the security of $\Pi$ with probability $\leq pab/(q-pab)$.
The upper bound can be made negligible in the statistical security parameter
$\lambda$ as long as   $q\approx 2^\lambda$ and $a,b,p $
are all polynomial functions in $\lambda$.
In our generic construction, the key $PK_F$ is  empty such that anyone, even without executing $\sf KeyGen$, is able to execute ${\sf ProbGen}(PK_F,{\bf x})$
to prepare its input vector $\bf x$.  Hence, the scheme $\Pi$ allows public delegation. One the other hand, the verification keys
 $VK_F, VK_{\bf x}$ must be kept private. Otherwise,
 an adversary will be able to easily to persuade the client to
 both accept a set of wrong server results and output a wrong function value.
Hence,    $\Pi$ is   privately verifiable.

For any  $F\in  \mathbb{Z}_q^{m\times d}$, the computational  cost of running ${\sf KeyGen}(\lambda, F)$
is dominated by $2amd$ additions modulo $q$ and  $amd$ multiplications   modulo $q$.
For any
${\bf  x}\in \mathbb{Z}_q^d$, the computational cost of running ${\sf ProbGen}(PK_F,{\bf  x})$ is dominated by
$bd$ additions modulo $q$.
For every $\ell\in [k]$, the  cost of
running   ${\sf Compute}(\ell,\rho_\ell, \sigma_\ell)$ at the $\ell$-th server  is
dominated by $|C_\ell|md$ additions modulo $q$ and $|C_\ell|md$ multiplications modulo $q$.
  The total cost of running all
$k$ server-side algorithms is dominated by  $\sum_{\ell=1}^k |C_\ell| md=abmd$ additions modulo $q$
and $\sum_{\ell=1}^k |C_\ell| md=abmd$  multiplications modulo $q$.
In verification, the client needs to compute
$ab$ inner products of dimension-$m$ vectors,  $ab$
inner product of dimension-$d$ vectors, and possibly $ab$ additions of dimension-$m$ vectors. The total computational  cost of executing
$\sf Verify$ is dominated
by $ab(2m+d)$ additions modulo $q$ and $ab(m+d)$ multiplications modulo $q$.
If we denote with ${\sf Add}_q$ additions modulo $q$ and
denote ${\sf Mul}_q$ multiplications modulo $q$, then the computational  cost of all algorithms in $\Pi$ can be summarized in
the following figure:
\begin{figure}[h]
\begin{center}
\begin{tabular}{|c|c|c|}
\hline
 Algorithms &   ${\sf Add}_q$
&    ${\sf Mul}_q$ \\
\hline
${\sf KeyGen}$ & $2amd$ & $amd$  \\ \hline
${\sf ProbGen}$ & $bd$& 0   \\ \hline
${\sf Compute}$ & $abmd$ & $abmd$  \\ \hline
${\sf Verify}$ & $ab(2m+d)$ & $ab(m+d)$   \\ \hline
\end{tabular}
\end{center}

\vspace{0mm}

\centering{Fig. 2: Computational Cost ($\Pi$)}
\label{fig:cs}
\end{figure}

The  client's total  computational cost  of  executing $\sf ProbGen$ and   $\sf Verify$ in the computing phase   is dominated by
$2abm+abd+bd$ additions modulo $q$ and
$ab(m+d)$ multiplications modulo $q$.
On the other hand, in the naive computation of $F{\bf x}$, the client has to do
 around $md$ additions modulo $q$ and
$md$ multiplications modulo $q$.
As the parameters $a,b$ are typically constants (see Section \ref{sec:opt} for parameter selection,
and Sections \ref{sec:3s} and \ref{sec:4s} for instantiations),
 we have that
$2abm+abd+bd=o(md)$ and $ab(m+d)=o(md)$ as long as  $m,d$ are large enough.
That is, the client's  total computational  cost in the computing phase of
 $\Pi$ will be substantially less than its cost in a naive computation of $F{\bf x}$.
  Therefore, our generic construction yields MSVC  schemes that are outsourceable.

\subsection{Parameter Selection and Optimization}
\label{sec:opt}

%

When  our generic construction $\Pi$ is instantiated,  the parameters $k,a,b, \\ \{A_\ell\}_{\ell=1}^k,
\{B_\ell\}_{\ell=1}^k$ and $\{C_\ell\}_{\ell=1}^k$ should be chosen  to both meet
 the requirements on input/function privacy and also optimize the
 computational cost.
In our MSVC model, the servers do not  communicate
with each other;  otherwise,  the input/function privacy may be compromised.
In an ideal instantiation, we prefer to choose a smallest  $k$ such that
 the number ${k\choose 2}$ of all possible 2-server collisions is minimized,
 in order to
guarantee the highest level of input/function privacy. On the other hand,
the client's total computational cost  in $\Pi$'s computing phase
 is roughly equal to
$2abm+abd+bd$ additions modulo $q$ and
$ab(m+d)$ multiplications modulo $q$; and the servers' total computational cost is roughly equal to
  $abmd$ additions modulo $q$ and
$abmd$ multiplications modulo $q$. For the fixed $m,d$, both workloads are minimized
as long as   $ab$ is minimized.
Therefore, in an ideal instantiation, we prefer to choose $a,b$ such that
$ab$ is  minimized,  in order to give the best  efficiency.
Minimizing both  $k$ and $ab$ simultaneously is infeasible, which will be seen soon.
In this section, we minimize  these parameters separately and obtain
two independent schemes: one attains the highest level of privacy,  and the other
 attains the best
efficiency.
\begin{definition}
\label{def:kab}
Let $k,a,b\geq 2$ be integers and let $A=[a], B=[b]$. The nonempty sets $A_1\times B_1,\ldots, A_k\times B_k$
form  a {\em $k$-covering} of $A\times B$ if
  $A_1,\ldots,A_k\subsetneq A$, $B_1,\ldots, B_k\subsetneq B$, and
  $(A\times B)\subseteq \bigcup_{\ell=1}^k (A_\ell\times B_\ell)$.
  For all $a,b\geq 2$, we define
{${k(a,b)}$} to be the smallest integer $k\geq 2$ such that there is a $k$-covering of
$[a]\times [b]$.
 For every integer $k\geq 2$, we define {${w(k)}$} to be the least
values of $ab$  such that  $[a]\times [b]$ has a $k$-covering, where $a,b\geq 2$.
We agree that $w(k)=\infty$ if for all $a,b\geq 2$ there is no $k$-covering  of
$[a]\times [b]$.
\end{definition}

\begin{theorem}
\label{thm:mk}
We have that $k(2,b)=k(a,2)=4$ and $k(a,b)=3$ for all $a,b\geq 3$.
\end{theorem}
\begin{proof}
For all $a,b\geq 2$, the sets
$\{1\}\times \{1\}, \{1\}\times \{2,\ldots,b\},  \{2,\ldots,a\}\times \{1\},$ and $
\{2,\ldots,a\}\times\{2,\ldots,b\}$  form a 4-covering of $[a]\times [b]$. Therefore,
$k(a,b)\leq 4$.
Due to Definition \ref{def:kab}, it is easy to see  that $k(a,b)\geq 2$ for all $a,b\geq 2$.
Hence, $k(a,b)\in \{2,3,4\}$ for all $a,b\geq 2$.

Below we show that $k(a,b)> 2$ for all $a,b\geq 2$. Assume for contradiction that there are integers
$a,b\geq 2$ such that $k(a,b)=2$. Then  there is a 2-covering $\{A_1\times B_1, A_2\times B_2\}$ of
$[a]\times [b]$, where the definition of covering shows that
$\emptyset \neq A_1,A_2\subsetneq [a], \emptyset \neq B_1, B_2\subsetneq [b]$ and $[a]\times [b]\subseteq (A_1\times B_1)\cup (A_2\times B_2)$.
Let $x\in [a]\setminus A_1$ and $y\in [b]\setminus B_2$. Then it is easy to see that
$(x,y)\in [a]\times [b], (x,y)\notin A_1\times B_1, (x,y)\notin A_2\times B_2$, which contradicts to
the requirement that
$[a]\times [b]\subseteq (A_1\times B_1)\cup (A_2\times B_2)$.

As a result, we have that $k(a,b)\in \{3,4\}$ for all $a,b\geq 2$.
Whenever $a,b\geq 3$, the sets $\{1,2\}\times \{1,2\},\{1,3,\ldots,a\}\times \{1,3,\ldots,b\},\{2,3\ldots,a\}\times \{2,3,\ldots, b\}$ would be a 3-covering of $[a]\times [b]$. Hence, we have that $k(a,b)=3$ for all $a,b\geq 3$.
At last, we show that $k(a,2)=4$. A similar proof for $k(2,b)=4$ exists and will be omitted from here.
Assume for contradiction that  $\{A_1\times B_1, A_2\times B_2, A_3\times B_3\}$
is a 3-covering of $[a]\times [2]$.
Due to the definition of covering, we have that $B_1,B_2,B_3\in
\{\{1\}, \{2\}\}$. Then at least one of $\{1\}$ and $\{2\}$ appears
at  most once in $B_1, B_2, B_3$.
Without loss of generality, suppose that $\{1\}$ appears at most once in $\{B_1,B_2,B_3\}$.
We distinguish between two cases: (i) $\{1\}$ does not appear in
$\{B_1,B_2,B_3\}$; (ii) $\{1\}$ appears once in $\{B_1,B_2,B_3\}$.
In the first case, $\{A_1\times B_1, A_2\times B_2, A_3\times B_3\}$ cannot form a covering
of $[a]\times [2]$ because for every $x\in [a]$,  $(x,1)\in [a]\times [2]$ but
$(x,1)\notin  (A_1\times B_1)\cup (A_2\times B_2)\cup (A_3\times B_3)$.
In the second case, we suppose that $B_1=\{1\}$ and $ B_2=B_3=\{2\}$.
As $A_1$ is a proper subset of $[a]$, we can choose $x\in [a]\setminus A_1$.
Then we would have $(x,1)\in [a]\times [2]$, but
$(x,1)\notin  (A_1\times B_1)\cup (A_2\times B_2)\cup (A_3\times B_3)$, which shows a contradiction.
Hence, $[a]\times [2]$ cannot have a 3-covering and  $k(a,2)$ must be equal to 4.
\end{proof}

\begin{theorem}
\label{thm:mab}
We have that $w(2)=\infty, w(3)=9$ and $w(k)=4$ for all $k\geq 4$.
\end{theorem}

\begin{proof}
The proof of Theorem \ref{thm:mk} shows that there is no 2-covering of
$[a]\times [b]$ for all $a,b\geq 2$. Hence, $w(2)=\infty$.
There is a 3-covering of $[a]\times [b]$ if and only if $a,b\geq 3$.
Among the choices of $a,b$, the product $ab$
is minimized when $a=b=3$. Therefore, we have that $w(3)=9$.
When $k\geq 4$, there is a $k$-covering of $[a]\times [b]$
for all $a,b\geq 2$.
Among the choices of $a,b$, the product $ab$
is minimized when $a=b=2$. Therefore, we have that $w(k)=4$ for all $k\geq 4$.
\end{proof}

\subsection{Instantiation with the Least Number of Servers}
\label{sec:3s}

Theorem \ref{thm:mk}  shows that   the smallest number of
required servers is 3 when the generic construction $\Pi$ is instantiated.
Let
$k=3, a=b=3,  A=[3], B=[3], A_1=\{1,2\}, B_1=\{1,2\},A_2=\{1,3\}, B_2=\{1,3\},A_3=\{2,3\}$ and $B_3=\{2,3\} $.
Then it is easy to verify that $A_1\times B_1, A_2\times B_2$ and $ A_3\times B_3$ form a cover
of $A\times B$ and  $C_1=\{(1,1),(1,2),(2,1),(2,2)\}\subseteq A_1\times B_1, C_2=\{(1,3),(3,1),(3,3)\}$
and $C_3=\{(2,3),(3,2)\}$ form a partition of $A\times B$.
By instantiating the generic construction $\Pi$ with the
 parameters $(k,a,b,A,B,\{A_\ell\}_{\ell=1}^3,\{B_\ell\}_{\ell=1}^3,\{C_\ell\}_{\ell=1}^3)$, we will get   3-server verifiable computation scheme, denoted as
 $\Pi_{\rm s}$, which requires  the smallest number of servers
 among all possible instantiations of $\Pi$.
Based on Figure 2, the computational cost of the algorithms in
$\Pi_{\rm s}$  can be summarized with  Fig. 3.
 \begin{figure}[H]
\begin{center}
\begin{tabular}{|c|c|c|}
\hline
 Algorithms & ${\sf Add}_q$
&  ${\sf Mul}_q$  \\
\hline
${\sf KeyGen}$ & $6md$ & $3md$  \\ \hline
${\sf ProbGen}$ & $3d$& 0   \\ \hline
${\sf Compute}$ & $9md$ & $9md$  \\ \hline
${\sf Verify}$ & $9(2m+d)$ & $9(m+d)$   \\ \hline
\end{tabular}
\end{center}
\centering{Fig. 3: Computational Cost ($\Pi_{\rm s}$)}
\end{figure}

\subsection{Instantiation with the Least Workload}
\label{sec:4s}

Theorem \ref{thm:mab}  shows that   the smallest amount of client/server computation will be done
 when the generic construction $\Pi$ is instantiated
to as a 4-server  scheme.
Let
$k=4, a=b=2,   A=[2], B=[2], A_1=\{1\}, B_1=\{1\},A_2=\{1\}, B_2=\{2\},A_3=\{2\}, B_3=\{1\}, A_4=\{2\}$ and $B_4=\{2\} $.
Then it is easy to verify that $A_1\times B_1, A_2\times B_2$,  $ A_3\times B_3$
and $A_4\times B_4$ form a cover
of $A\times B$ and  $C_1=\{(1,1) \}\subseteq A_1\times B_1, C_2=\{(1,2)\}$,
$C_3=\{(2,1)\}$ and $C_4=\{(2,2)\}\subseteq A_4\times B_4$ form a partition of $A\times B$.
By instantiating the generic construction $\Pi$ with the
 parameters $(k,a,b,A,B,\{A_\ell\}_{\ell=1}^3,\{B_\ell\}_{\ell=1}^3,\{C_\ell\}_{\ell=1}^3)$,
  we get a  4-server verifiable computation scheme, denoted as
 $\Pi_{\rm w}$, which has the fastest client/server computation among all possible instantiations of $\Pi$.
Based on Figure 2, the computational cost of the algorithms in
$\Pi_{\rm w}$ can be summarized with Fig. 4.
\begin{figure}[H]
\begin{center}
\begin{tabular}{|c|c|c|}
\hline
 Algorithms &  ${\sf Add}_q$
&  ${\sf Mul}_q$  \\
\hline
${\sf KeyGen}$ & $4md$ & $2md$  \\ \hline
${\sf ProbGen}$ & $2d$& 0   \\ \hline
${\sf Compute}$ & $4md$ & $4md$  \\ \hline
${\sf Verify}$ & $4(2m+d)$ & $4(m+d)$   \\ \hline
\end{tabular}
\end{center}

\vspace{0mm}

\centering{Fig. 4: Computational Cost ($\Pi_{\rm w}$)}

\end{figure}

\subsection{Implementation}

Our generic construction is practical and suitable for
implementation.
In order to show its practicality, we have implemented
  $\Pi_{\sf s}$ and  $\Pi_{\sf w}$ on a DELL Precision Tower T7810 workstation
that runs with  the  Intel  Xeon   E5-2650   (2.30 GHz) Processor.
In both cases, we implemented all algorithm   on the same platform, in order to
compare between the computational costs.
In our implementations, we  set
 $q=82434016654300709346097073375351854135999471015108634126889281238621513052057
$
 to be a 256-bit   prime such that the success probability of any adversary
 that makes $p$ attempts in breaking $\Pi_{\rm s}$ and $\Pi_{\rm w}$ is bounded by $\frac{9p}{2^{256}-9p}$ and
$\frac{4p}{2^{256}-4p}$ respectively.
\begin{figure}[H]
\begin{center}
\begin{minipage}{8cm}
\begin{center}
\includegraphics[scale=0.395]{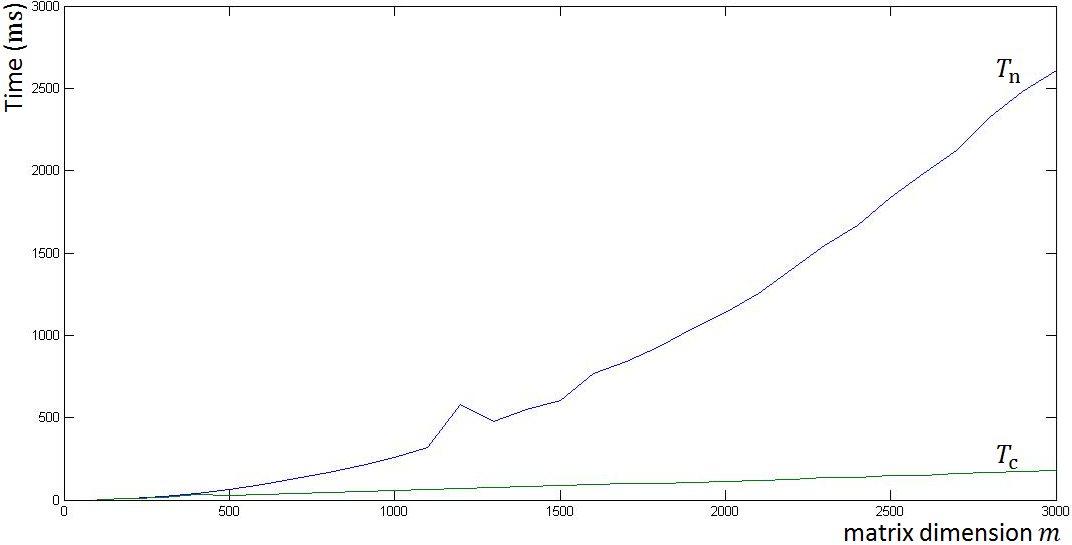}

\vspace{0mm}
{ \hspace{1cm}   Fig 5: Client Computation ($\Pi_{\rm s}$)}
\end{center}
\end{minipage}
\hspace{5mm}
\begin{minipage}{8cm}
\begin{center}
\includegraphics[scale=0.394]{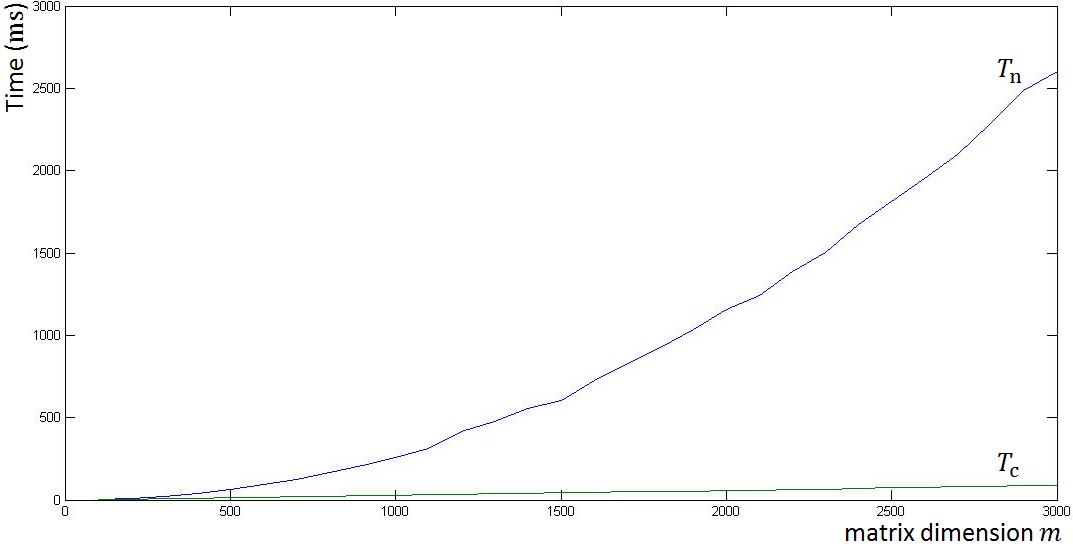}

\vspace{0mm}
{ \hspace{1cm}  Fig 6:  Client Computation ($\Pi_{\rm w}$)}
\end{center}
\end{minipage}
\end{center}
\end{figure}

In our implementations, we choose $m=d$ and consider both  a random   function $F\in \mathbb{Z}_q^{m\times d}$ and a
random  input ${\bf x}\in \mathbb{Z}_q^{d}$.
We record both the time $T_{\rm n}$ required by the naive computation of
   $F{\bf x}$
and the total time $T_{\rm c}$ required by the  client's execution of
$\sf ProbGen$ and $\sf Verify$ in the computing phase.
In both experiments we get the results for  $m\in \{100,200,\ldots,3000\}$, which are shown
in Fig. 5 and Fig. 6, respectively.
Our experiments show that $T_{\rm c}$ is much smaller than $T_{\rm n}$, i.e.,
 the client computation  in $\Pi_{\rm s}$ and $\Pi_{\rm w}$
are substantially faster than the naive computation of
$F{\bf x}$.
For example, when
$m=d=3000$, we have that
$(T_{\rm n}, T_{\rm c})={\rm (2609.08ms,  176.74ms)}$ in $\Pi_{\rm s}$
and
$(T_{\rm n}, T_{\rm c})={\rm (2801.35,  89.77ms)}$ in $\Pi_{\rm w}$.
In the literature of SSVC,
few schemes allow  us to observe that
$T_{\rm c}<T_{\rm n}$,
due to the use of expensive operations such as FHE, GCs, or heavy
public-key operations.

 \section{Applications}

\subsection{Analysis of Sensitive Data}

In some applications the function  $F$ may be a
trading secret that can be  expressed as a matrix function or a  polynomial function.
The developer of this algorithm    wants to provide paid services by
 applying the algorithm to the user's sensitive data.
   However, the developer may not be   willing to maintain his own computing
    infrastructures, which will incur significant financial cost.
In this scenario, the developer can outsource the computation of
$F$ to multiple clouds  with our MSVC schemes such that: (1) both the algorithm
$F$ and the user's data are kept secret from the servers;
(2) the servers' results are verifiable in order to
guarantee the correct reconstruction of output; (3)
the result verification and reconstruction  are substantially faster than
the naive computation of $F$.

For example, in molecular anthropology the single nucleotide polymorphisms (SNPs)
are usually analyzed in order to help diagnosis, give medication and
conduct research. The human genome may be a sequence of
$\{\rm A,C,G,T\}$ of length $\approx 3\times 10^9$.
Genomes of two individuals are $99.6\%$ similar and part of the differences is
contributed by SNPs. Many SNPs explain why some people have higher chance to
have diabetes, cancers, and other inherited diseases than others.
It is well-known that how likely a patient will have a disease can be computed from his
SNPs.
Let $\{{\sf SNP}_1, {\sf SNP}_2,\ldots, {\sf SNP}_d\}$ be a set of
related SNPs    when a set $\{{\sf D}_1,{\sf D}_2,\ldots, {\sf D}_m\}$
of diseases are considered.
A vector ${\bf x}=({\bf x}[1],{\bf x}[2],\ldots, {\bf x}[d])\in \{0,1,2\}^d$ can be defined such that for every $j\in[d]$,
${\bf x}[j]$ stands for  the number of times that ${\sf SNP}_j$ occurs in
a person's genome.  An $m\times d$ matrix $F=(F[i,j])_{m\times d}$ can be defined such that
for every $i\in[m]$, the $i$-th row of $F$  is a set of important factors for the disease
 ${\sf D}_i$ and the risk of a person
 to have disease ${\sf D}_i$ can be computed as
$\sum_{j=1}^m F[i,j]{\bf x}[j].$ Then the risk
of the person to have the diseases $\{{\sf D}_1,{\sf D}_2,\ldots, {\sf D}_m\}$
can be computed as $F{\bf x}$.
In this example, the developer (e.g., a doctor) of the algorithm  may
have spent significant research  efforts in developing  $F$ and thus consider
$F$ as a trading secret; any user (e.g.,  a patient) of the algorithm $F$ must keep his data
$\bf x$ secret as well.
Our MSVC schemes would allow the doctor to outsource the computation of $F{\bf x}$
to multiple clouds, such that (1)-(3) are all satisfied.

\subsection{Polynomial Outsourcing}

Achieving
input privacy in SSVC has been
highly non-trivial, especially when the   function $F$ requires
high degree computations on the input $x$.
On one hand, in order to keep
$x$ private from the server, the client  has to encrypt $x$
as ${\sf Enc}(x)$ with a semantically secure encryption   $\sf Enc$.
On the other hand, the server has to compute
$F(x)$ with $F$ and ${\sf Enc}(x)$, which   requires
 $\sf Enc$ to be    homomorphic.
 However, the homomorphic encryptions are still far from  practical \cite{MSM18} today.
  As a result,  the   resulting SSVC  schemes  are usually
impractical, in terms of the client-side computation and the server-side computation.

Our MSVC schemes in Section \ref{sec:con}  enable  the computation of
$F  {\bf  x}$ for any   $F\in \mathbb{Z}_q^{m\times d}$ and
${\bf  x}\in \mathbb{Z}_q^d$; allow public delegation  and private verification;
have information-theoretic security and  privacy (of input and function);
    and are practically efficient. They successfully resolved the conflict  between
input privacy and practicality.
In this section, we shall show  how to use these schemes to resolve the same conflict
in the delegation of various polynomial functions, which may have very high degrees.
Our main observation is that the  special algebraic structure  of a polynomial $f$ may   allow us to decompose the computation of
$f(x)$ as a  {\em two-stage computation}, where  the {\em first-stage} computation is
a matrix-vector multiplication of the form ${\bf u}=F{\bf x}$ and the {\em second-stage} computation
is an inner product computation ${\bf y}\cdot {\bf u}$.
In particular, $F$ is a matrix determined by $f$ and $\bf x,y$ are vectors determined by
$x$. Usually   the computation of $F{\bf x}$ is heavy and that of $\bf y\cdot u$ is fast. By delegating the heavy computation of $F{\bf x}$ with our MSVC scheme and
leaving the light computation of $\bf y\cdot u$ to the client, the client would be able to offload the main workload to the servers.
Furthermore, the resulting delegation scheme for $f$ would inherit all
nice properties of MSVC, such as correctness, security, input/function privacy,
and practical efficiency.
The method of decomposing polynomial evaluations into two-stage computations will be detailed as below.
\begin{itemize}
\item
{\bf Univariate polynomials.}
For any univariate polynomial   $f(x)=\sum_{i=0}^d f_ix^i$, define   $m=\lceil \sqrt{d+1}\rceil$;
\begin{equation*}
F=
\left(
  \begin{array}{cccc}
    f_{0} & f_{1} & \cdots & f_{m-1} \\
    f_{m} & f_{m+1} & \cdots & f_{2m-1} \\
    \vdots & \vdots & \cdots & \vdots \\
    f_{m^2-m} & f_{m^2-m+1} & \cdots & f_{m^2-1} \\
  \end{array}
\right),
\end{equation*}
 where
$f_{i}=0$ for all $i>d$;
  ${\bf  x}=(1,x,\ldots,x^{m-1})^\top$; and  ${\bf  y}=(1,x^m, \ldots, x^{m^2-m})$.
   Then we will have that $f(x)={\bf y}(F{\bf x})$.
The first-stage computation   requires
$O(m^2)=O(d)$ arithmetic operations modulo $q$, and the second-stage computation
  requires
$O(m)=O(\sqrt{d})$ arithmetic operations modulo $q$.

\item
{\bf Bivariate polynomials.}
For any  bivariate polynomial\\
$f(x,y)=\sum_{i,j=0}^d f_{i,j}x^i y^j$,
define  $F=(f_{i,j})$,
   ${\bf  x}=(1,x,\ldots,x^d)^\top$ and
${\bf  y}=(1,y,\ldots,y^d)$.
Then we will  have that
$f(x,y)={\bf  y}( F  {\bf  x})$.
The first-stage computation   requires
$O(d^2)$ arithmetic operations modulo $q$, and the second-stage computation  requires
$O(d)$ arithmetic operations modulo $q$.

\item
{\bf Quadratic multivariate polynomials.}
For a   polynomial
$f(x_1,\ldots, x_d)=\sum_{i,j=1}^d f_{i,j}\cdot x_i x_j$, define
$F=(f_{i,j})$,
  ${\bf  x}=(x_1,\ldots,x_d)^\top$ and ${\bf y}={\bf x}^\top$.
   Then we will    have that
$f(x_1,\ldots,x_d)={\bf  y} (F{\bf  x})$.
The first-stage computation  requires
$O(d^2)$ arithmetic  operations modulo $q$, and the second-stage computation   requires
$O(d)$ arithmetic  operations modulo $q$.

\item
{\bf Multivariate polynomials of bounded degree in each variable.}
For a polynomial
$f(x_1,\ldots, x_m)=\sum_{i_1,\ldots,i_m=1}^d f_{i_1,\ldots,i_m}\cdot x_1^{i_1}\cdots x_m^{i_m}$, define  $\ell=\lfloor m/2 \rfloor$,
 ${\bf  y}=(x_1^{i_1}\cdots x_\ell^{i_\ell})\in \mathbb{Z}_q^{d^l}$,
 ${\bf  x}=(x_{\ell+1}^{i_{\ell+1}}\cdots x_m^{i_m})^\top\in \mathbb{Z}_q^{d^{m-l}}$; and let
$
F=\big(F_{(i_1,\ldots,i_\ell),(i_{\ell+1},\ldots,i_m)}\big)
$
be a matrix of size  $d^\ell\times d^{m-\ell}$ such that
$F_{(i_1,\ldots,i_\ell),(i_{\ell+1},\ldots,i_m)}=
f_{i_1,\ldots,i_m}$ for all $(i_1,\ldots,i_\ell)\in [d]^\ell$ and
$(i_{\ell+1},\ldots, i_m)\in [d]^{m-\ell}$.
Then we will  have that $f(x_1,\ldots,x_m)={\bf  y} (F {\bf  x})$.
The first-stage computation  requires
$O(d^{m})$ arithmetic  operations modulo $q$, and the second-stage computation  requires
$O(d^{\lfloor m/2 \rfloor})$  arithmetic operations modulo $q$.
\end{itemize}

In all the above decompositions,
 the first-stage computation is as heavy as the naive computation of $f(x)$ and
 the second-stage computation is substantially faster. By delegating the first-stage computations with our MSVC schemes, we would obtain
the expected  MSVC schemes for polynomials.

\subsection{Protocol Design}

Our MSVC schemes   have interesting applications
in the design of  cryptographic  protocols such as outsourced private information retrieval.
A $t$-private $k$-server information-theoretic private information retrieval (PIR)
 \cite{CGKS95} is a   protocol between a client and $k$ servers, where each server has a database
$f=(f_1,f_2,\ldots,f_N)$ and the client is interested in
$f_i$ for some $i\in[N]$. Such a protocol allows the client to
retrieve $f_i$ from the servers such that  $i$ is still   hidden from any $\leq t$ servers.
Its efficiency  is mainly measured with two parameters:
(1)
 {\em communication complexity},  which is  the total number of bits communicated for retrieving
a single entry of $f$;
(2) {\em server computation complexity}: which is
the total number of database entries  accessed by the servers in each retrieval.

While the communication complexity can be sublinear in $N$, Beimel et al.
 \cite{BIM00} has shown that the servers' computation complexity
 is $\Omega(N)$ for any PIR.
When $N$ is large,  it would be nice to   outsource   PIR servers' computation to cloud  services \cite{HG13,MBC13}, which have numerous computing resources.
 However, outsourcing requires
 a stronger  adversarial  model as the cloud may be  untrusted.
Most of the existing PIR protocols \cite{KO97}   assume that the PIR servers should be honest-but-curious.
In the cloud scenario,  this weak model  should be strengthened  to resist  malicious servers. A natural way is by making the PIR server's computation   verifiable.

In the literature, the specific computations studied in this paper,  such as
matrix-vector multiplications and polynomial evaluations, have been widely used in  PIR construction.
For example, one can consider   $f$
as the entries  of a square matrix
\begin{equation}
F=(F_{i,j})=
\left(
  \begin{array}{cccc}
    f_{1} & f_{2} & \cdots & f_{d} \\
    f_{d+1} & f_{d+2} & \cdots & f_{2d} \\
    \vdots & \vdots & \cdots & \vdots \\
    f_{d^2-d+1} & f_{d^2-d+2} & \cdots & f_{d^2} \\
  \end{array}
\right),
\end{equation}
where $d=\lceil \sqrt{N}\rceil$ and $f_j=0$ for all $j>N$.
When the client is interested in $f_i$ and
 $f_i$ is  the $(r,c)$-entry of $F$ for some $r,c\in[d]$,
it suffices for the client to privately retrieve the
$c$-th column of $F$, i.e., $(F_{1,c}, \ldots, F_{d,c})$.
Then the server-side computation can be captured by
 $F{\bf  x}$ for ${\bf  x}=(0,\ldots,1,\ldots,0)^\top$, where the $c$th component of $\bf x$ is equal to 1 and all  other components
are equal to 0.
Our MSVC schemes    would enable the client to delegate the computation of
$F  {\bf  x}$ to the PIR cloud servers such that:
 (1)  ${\bf  x}$ is kept secret from each   server;
 (2) the server's computation results become verifiable.

In this  simple solution, the client and the server  only need to communicate
$O(\sqrt{N})$  bits, which gives a nontrivial outsourced  PIR.
As an additional property, the database $f$   is information-theoretically hidden from
each  server.

\section{Conclusion}

In this paper, we defined an MSVC model where
each server performs a partial computation on the function shares and input shares.
We give a generic construction that achieves public delegation and  private verification,
information-theoretic security, input and function privacy.
We also apply these schemes to construct MSVC schemes for various polynomial functions.
 Our schemes are  free of public-key operations and  practically efficient.
Our schemes also yield   information-theoretic PIR schemes that are secure againat
malicious servers.
We leave  it as a  future work to design MSVC schemes with public verification.

\section*{Acknowledgement}

The research was supported by Singapore Ministry of Education under Research Grant RG12/19 and National Natural
Science Foundation of China (No. 61602304). The author would like to thank the anonymous referees for the helpful comments.


\begin{thebibliography}{10}

\bibitem{ACG+14}
P.~Ananth, N.~Chandran, V.~Goyal, B.~Kanukurthi, and R.~Ostrovsky.
\newblock Achieving privacy in verifiable computation with multiple
  servers–without {FHE} and without pre-processing.
\newblock {\em in: Proceedings of the 17th International Conference on
  Public-Key Cryptography, PKC, Springer}, pages 149--166, 2014.

\bibitem{AIK10}
B.~Applebaum, Y.~Ishai, and E.~Kushilevitz.
\newblock From secrecy to soundness: efficient verification via secure
  computation.
\newblock {\em in: Proceedings of the 37th International Colloquium on
  Automata, Languages, and Programming, ICALP, Springer}, pages 152--163, 2010.

\bibitem{Bab85}
L.~Babai.
\newblock Trading group theory for randomness.
\newblock {\em in: Proceedings of the 17th Annual ACM symposium on Theory of
  computing, STOC, ACM}, pages 421--429, 1985.

\bibitem{BF12}
M.~Barbosa and P.~Farshim.
\newblock Delegatable homomorphic encryption with applications to secure
  outsourcing of computation.
\newblock {\em in: Proceedings of The Cryptographers’ Track at the RSA
  Conference 2012, CT-RSA, Springer}, pages 296--312, 2012.

\bibitem{Bei96}
A.~Beimel.
\newblock Secure schemes for secret sharing and key distribution.
\newblock {\em Ph.D. thesis, Israel Institute of Technology}, 1996.

\bibitem{BIM00}
A.~Beimel, Y.~Ishai, and T.G. Malkin.
\newblock Reducing the servers computation in private information retrieval:
  {PIR} with preprocessing.
\newblock {\em in: Proceedings of the 20th Annual International Cryptology
  Conference, CRYPTO, Springer}, pages 55--73, 2000.

\bibitem{BGKW88}
M.~Ben-Or, S.~Goldwasser, J.~Kilian, and A.~Wigderson.
\newblock Multi-prover interactive proofs: how to remove intractability
  assumptions.
\newblock {\em in: Proceedings of the 20th Annual ACM Symposium on Theory of
  Computing, STOC, ACM}, pages 113--131, 1988.

\bibitem{BGV11}
S.~Benabbas, R.~Gennaro, and Y.~Vahlis.
\newblock Verifiable delegation of computation over large datasets.
\newblock {\em in: Proceedings of the 31st Annual Cryptology Conference,
  CRYPTO, Springer}, pages 111--131, 2011.

\bibitem{snark}
N.~Bitansky, R.~Canetti, A.~Chiesa, and E.~Tromer.
\newblock From extractable collision resistance to succinct noninteractive
  arguments of knowledge, and back again.
\newblock {\em in: Proceedings of the 3rd Innovations in Theoretical Computer
  Science Conference, ITCS, ACM}, pages 326--349, 2012.

\bibitem{BLR90}
M.~Blum, M.~Luby, and R.~Rubinfeld.
\newblock Self-testing/correcting with applications to numerical problems.
\newblock {\em in: Proceedings of the 22nd Annual ACM symposium on Theory of
  Computing, STOC, ACM}, pages 73--83, 1990.

\bibitem{BW94}
M.~Blum and H.~Wasserman.
\newblock Program result-checking: a theory of testing meets a test of theory.
\newblock {\em in: Proceedings of the 35th Annual Symposium on Foundations of
  Computer Science, FOCS, IEEE}, pages 382--392, 1994.

\bibitem{BV11}
Z.~Brakerski and V.~Vaikuntanathan.
\newblock Fully homomorphic encryption from {Ring-LWE} and security for key
  dependent messages.
\newblock {\em in: Proceedings of the 31st Annual Cryptology Conference,
  CRYPTO, Springer}, pages 505--524, 2011.

\bibitem{CRR112}
R.~Canetti, B.~Riva, and G.~N. Rothblum.
\newblock Practical delegation of computation using multiple servers.
\newblock {\em in: Proceedings of the 18th ACM Conference on Computer and
  Communications Security, CCS, ACM}, pages 445--454, 2011.

\bibitem{CRR11}
R.~Canetti, R.~Riva, and G.~Rothblum.
\newblock Two protocols for delegation of computation.
\newblock {\em in: Proceedings of the 6th International Conference Information
  Theoretic Security, ICITS, Springer}, pages 37--61, 2012.

\bibitem{mac}
D.~Catalano and D.~Fiore.
\newblock Practical homomorphic {MACs} for arithmetic circuits.
\newblock {\em in: Proceedings of the 32nd Annual International Conference on
  the Theory and Applications of Cryptographic Techniques, EUROCRYPT,
  Springer}, pages 336--352, 2013.

\bibitem{FG13}
D.~Catalano, D.~Fiore, R.~Gennaro, and K.~Vamvourellis.
\newblock Algebraic (trapdoor) one-way functions and their applications.
\newblock {\em in: Proceedings of the 10th Theory of Cryptography Conference,
  TCC, Springer}, pages 680--699, 2013.

\bibitem{CGKS95}
B.~Chor, O.~Goldreich, E.~Kushilevitz, and M.~Sudan.
\newblock Private information retrieval.
\newblock {\em in: Proceedings of the 36th Annual Symposium on Foundations of
  Computer Science, FOCS, IEEE}, pages 41--50, 1995.

\bibitem{CKLR11}
K.M. Chung, Y.T. Kalai, F.H. Liu, and R.~Raz.
\newblock Memory delegation.
\newblock {\em in: Proceedings of the 31st Annual Cryptology Conference,
  CRYPTO, Springer}, pages 151--168, 2011.

\bibitem{CKV11}
K.M. Chung, Y.T. Kalai, and S.P. Vadhan.
\newblock Improved delegation of computation using fully homomorphic
  encryption.
\newblock {\em in: Proceedings of the 30th Annual Cryptology Conference,
  CRYPTO, Springer}, pages 483--501, 2010.

\bibitem{EOAM16}
K.~Elkhiyaoui, M.~{\"{O}}nen, M.~Azraoui, and R.~Molva.
\newblock Efficient techniques for publicly verifiable delegation of
  computation.
\newblock {\em in: Proceedings of the 11th ACM on Asia Conference on Computer
  and Communications Security, AsiaCCS, ACM}, pages 119--128, 2016.

\bibitem{FG12}
D.~Fiore and R.~Gennaro.
\newblock Publicly verifiable delegation of large polynomials and matrix
  computations, with applications.
\newblock {\em in: Proceedings of the 19th ACM Conference on Computer and
  Communications Security, CCS, ACM}, pages 501--512, 2012.

\bibitem{FGP14}
D.~Fiore, R.~Gennaro, and V.~Pastro.
\newblock Efficiently verifiable computation on encrypted data.
\newblock {\em in: Proceedings of the 21st ACM Conference on Computer and
  Communications Security, CCS, ACM}, pages 844--855, 2014.

\bibitem{GGP10}
R.~Gennaro, C.~Gentry, and B.~Parno.
\newblock Non-interactive verifiable computing: outsourcing computation to
  untrusted workers.
\newblock {\em in: Proceedings of the 30th Annual Cryptology Conference,
  CRYPTO, Springer}, pages 465--482, 2010.

\bibitem{qsp}
R.~Gennaro, C.~Gentry, B.~Parno, and M.~Raykova.
\newblock Quadratic span programs and succinct nizks without pcps.
\newblock {\em in: Proceedings of the 32nd Annual International Conference on
  the Theory and Applications of Cryptographic Techniques, EUROCRYPT,
  Springer}, pages 626--645, 2013.

\bibitem{fhe}
C.~Gentry.
\newblock Fully homomorphic encryption using ideal lattices.
\newblock {\em in: Proceedings of the 41st Annual ACM Symposium on Theory of
  Computing, STOC, ACM}, pages 169--178, 2009.

\bibitem{GKR08}
S.~Goldwasser, Y.T. Kalai, and G.N. Rothblum.
\newblock Delegating computation: interactive proofs for muggles.
\newblock {\em in: Proceedings of the 40th Annual ACM Symposium on Theory of
  Computing, STOC, ACM}, pages 113--122, 2008.

\bibitem{GMR85}
S.~Goldwasser, S.~Micali, and C.~Rackoff.
\newblock The knowledge complexity of interactive proof systems.
\newblock {\em SIAM J. Comput.}, 18(1):186--208, 1989.

\bibitem{sig}
S.~Gorbunov, V.~Vaikuntanathan, and D.~Wichs.
\newblock Leveled fully homomorphic signatures from standard lattices.
\newblock {\em in: Proceedings of the 47th Annual ACM symposium on Theory of
  Computing, STOC, ACM}, pages 469--477, 2015.

\bibitem{HG13}
Y.~Huang and I.~Goldberg.
\newblock Outsourced private information retrieval.
\newblock {\em in: Proceedings of the 12th ACM Workshop on Privacy in the
  Electronic Society, WPES, ACM}, pages 119--130, 2013.

\bibitem{JY14}
C.~Joo and A.~Yun.
\newblock Homomorphic authenticated encryption secure against chosen-ciphertext
  attack.
\newblock {\em in: Proceedings of the 20th International Conference on the
  Theory and Application of Cryptology and Information Security, ASIACRYPT,
  Springer}, pages 173--192, 2014.

\bibitem{Kil92}
J.~Kilian.
\newblock A note on efficient zero-knowledge proofs and arguments.
\newblock {\em in: Proceedings of the 24th Annual ACM symposium on Theory of
  Computing, STOC, ACM}, pages 723--732, 1992.

\bibitem{Kil95}
J.~Kilian.
\newblock Improved efficient arguments.
\newblock {\em in: Proceedings of the 15th Annual International Cryptology
  Conference, CRYPTO, Springer}, pages 311--324, 1995.

\bibitem{KO97}
E.~Kushilevitz and R.~Ostrovsky.
\newblock Replication is not needed: single database, computationally-private
  information retrieval.
\newblock {\em in: Proceedings of the 38th Annual Symposium on Foundations of
  Computer Science, FOCS, IEEE}, pages 363--373, 1997.

\bibitem{LPJY13}
B.~Libert, T.~Peters, M.~Joye, and M.~Yung.
\newblock Linearly homomorphic structure-preserving signatures and their
  applications.
\newblock {\em in: Proceedings of the 33rd Annual Cryptology Conference,
  CRYPTO, Springer}, pages 289--307, 2013.

\bibitem{MSM18}
P.~Martins, L.~Sousa, and A.~Mariano.
\newblock A survey on fully homomorphic encryption: an engineering perspective.
\newblock {\em ACM Computing Survey}, 50(6):83:1--83:33, 2018.

\bibitem{MBC13}
T.~Mayberry, E.-O. Blass, and A.H. Chan.
\newblock {PIRMAP}: efficient private information retrieval for mapreduce.
\newblock {\em in: Proceedings of the 17th International Conference on
  Financial Cryptography and Data Security, FC, Springer}, pages 371--385,
  2013.

\bibitem{Mic94}
S.~Micali.
\newblock {CS} proofs.
\newblock {\em in: Proceedings of the 35th Annual Symposium on Foundations of
  Computer Science, FOCS, IEEE}, pages 436--453, 1994.

\bibitem{Moh11}
P.~Mohassel.
\newblock Efficient and secure delegation of linear algebra.
\newblock {\em in: IACR Cryptology ePrint Archive}, 2011(605), 2011.

\bibitem{PST13}
C.~Papamanthou, E.~Shi, and R.~Tamassia.
\newblock Signatures of correct computation.
\newblock {\em in: Proceedings of the 10th Theory of Cryptography Conference,
  TCC, Springer}, pages 222--242, 2013.

\bibitem{PHGR13}
B.~Parno, J.~Howell, C.~Gentry, and M.~Raykova.
\newblock Pinocchio: nearly practical verifiable computation.
\newblock {\em in: Proceedings of the 34th IEEE Symposium on Security and
  Privacy, S\&P, IEEE}, pages 238--252, 2013.

\bibitem{PRV12}
B.~Parno, M.~Raykova, , and V.~Vaikuntanathan.
\newblock How to delegate and verify in public: verifiable computation from
  attribute-based encryption.
\newblock {\em in: Proceedings of the 9th Theory of Cryptography Conference,
  TCC, Springer}, pages 422--439, 2012.

\bibitem{Tha13}
J.~Thaler.
\newblock Time-optimal interactive proofs for circuit evaluation.
\newblock {\em in: Proceedings of the 33rd Annual Cryptology Conference,
  CRYPTO, Springer}, pages 71--89, 2013.

\end{thebibliography}
\end{document}